\documentclass[aps,prb,twocolumn,showpacs,superscriptaddress]{revtex4-1}
\usepackage[colorlinks ,linkcolor=blue,anchorcolor=blue,citecolor=blue,urlcolor=blue]{hyperref}
\usepackage{graphicx}
\usepackage{epstopdf}
\def\avg#1{\langle#1\rangle}

\def\be{\begin{equation}} \def\ee{\end{equation}}
\def\bea{\begin{eqnarray}} \def\eea{\end{eqnarray}}

\def\nn{\nonumber}

\begin{document}

\title{Mott insulating states and quantum phase transitions of
correlated SU($2N$) Dirac fermions}
\author{Zhichao Zhou}
\affiliation{School of Physics and Technology, Wuhan University, Wuhan
430072, China}
\author{Da Wang}
\affiliation{National Laboratory of Solid State Microstructures and
 School of Physics, Nanjing University, Nanjing, 210093, China}
\author{Zi Yang Meng}
\affiliation{Beijing National Laboratory for Condensed Matter Physics, and Institute of Physics, Chinese Academy of Sciences, Beijing 100190, China}
\author{Yu Wang}
\email{yu.wang@whu.edu.cn}
\affiliation{School of Physics and Technology, Wuhan University, Wuhan
430072, China}
\author{Congjun Wu}
\email{wucj@physics.ucsd.edu}
\affiliation{Department of Physics, University of California, San Diego,
CA 92093, USA}

\begin{abstract}
The interplay between charge and spin degrees of freedom in strongly
correlated fermionic systems, in particular of Dirac fermions, is
a long-standing problem in condensed matter physics.
We investigate the competing orders in the half-filled SU($2N$) Hubbard
model on a honeycomb lattice, which can be accurately realized
in optical lattices with large-spin ultra-cold alkaline-earth fermions.
Employing large-scale projector determinant quantum Monte Carlo simulations,
we have explored quantum phase transitions from the gapless Dirac
semi-metals to the gapped Mott-insulating phases in the SU($4$) and SU($6$)
cases.
Both of these Mott-insulating states are found to be columnar valence bond
solid (cVBS) and to be absent of the antiferromagnetic N\'{e}el ordering
and the loop current ordering.
Inside the cVBS phases, the dimer ordering is enhanced by increasing
fermion components and behaves non-monotonically as the
interaction strength increases.
Although the transitions generally should be of first order
due to a cubic invariance possessed by the cVBS order,
the coupling to gapless Dirac fermions can soften
the transitions to second order through a non-analytic term in
the free energy.
Our simulations provide important guidance for the experimental
explorations of novel states of matter with ultra-cold alkaline
earth fermions.

\end{abstract}
\maketitle

\section{Introduction}
Massless Dirac fermions, describing the low energy quasiparticles
on the honeycomb lattice, have received a great deal of attention
in the past decade due to the rapid progress in graphene-related
and topological materials \cite{CastroNeto2009,Kotov2012,Na3Bi2014,Cd3As22014}.
On the theoretical aspect, how the gapless fermions develop gap
is an important scientific question both to condensed matter
and high-energy physics \cite{Ryu2009,Venkitesh2015}.
In condensed matter systems, gap opening is often accompanied by
symmetry breaking developing charge or spin orders.
On the other hand, interaction can also open energy gap in the Mott
insulating phase with or without spontaneous symmetry breaking.
The simplest theoretical model describing the correlated electrons is
the Hubbard model \cite{Hubbard1963,*Hubbard1964} which preserves the
SU($2$) symmetry.
The studies of the SU($2$) Hubbard model on the honeycomb lattice have
shown a quantum phase transition from the massless Dirac semi-metal
phase to a Mott insulating phase accompanied by the appearance
of the long-range antiferromagnetic N\'{e}el ordering, even though
whether these two transitions occur simultaneously is still
under debate \cite{Meng2010,Sorella2012,Assaad2013,Toldin2015}.

One important question is the possible novel physics if fermions
possess more than two internal components.
In this case, the SU(2) Hubbard model can be generalized to the
SU(2N) one \cite{Affleck1988,Lu1994,Honerkamp2004}.
(Since the fermion component number is typically even, we denote it
$2N$ throughout this article.)
This generalization is not only out of curiosity, but also can actually
be implemented in the state-of-art cold atom experiments with
large-spin alkaline-earth fermions
\cite{DeSalvo2010,Taie2010, Taie2012,Zhang2014,*scazza2014,*pagano2014,
Bloch2008,Jordens2008, Schneider2008}.
These large-spin fermions are fundamentally different from the
large-spin systems in solids which usually only exhibit the
SU(2) symmetry.
In contrast, the large-spin ultra-cold fermions provide a route
to realize high symmetries without fine tuning \cite{Wu2010a, *Wu2012}.
As an early effort, the simplest large spin fermion systems with
spin-$\frac{3}{2}$, including both alkali and alkaline-earth
fermions  were proved to be generically SO(5), or, isomorphically,
Sp(4) symmetric without fine tuning \cite{Wu2003,Wu2005,Hattori2005,
Controzzi2006,WU2006}.
For alkali-earth fermions, their interactions are spin-independent.
Therefore, such systems naturally realize the SU($2N$) symmetry
with $2N=2I+1$ and $I$ the fermion hyper-fine spin number,
as observed in several recent cold atom experiments.
Motivated by these theoretical and experimental progresses, the SU($2N$)
Hubbard model has provoked our interest and been studied systematically
in several recent works but only focusing on the square lattice
\cite{Cai2013,Cai2013a,Wang2014,Zhou2014}.
These studies show that the full understanding of the SU($2N$) Hubbard
models requires equal-footing treatments of both the small-$U$
Slater physics and the large-$U$ Mott physics.

The competition between the charge and spin degrees of freedom
of the SU($2N$) Hubbard model is a challenging problem due to
its non-perturbative nature.
At half-filling and in the large-$U$ limit, where charge fluctuations
are suppressed, the low energy spin degrees of freedom are described by
the SU($2N$) Heisenberg model with the super-exchange energy scale
$J=4t^2/U$ \cite{Affleck1985,Arovas1988}, {\it i.e.},
each site is under the constraint of $N$ fermions per site.
The large-$N$ expansion technique, in which $1/N$ is used as
a perturbative parameter, has been applied to study the
SU($2N$) Heisenberg model.
The VBS states, also named as the dimer ordering states,
were found  on both the square and honeycomb lattices for large enough
$N$ \cite{Read1989,*Read1989a}.
Recent quantum Monte Carlo (QMC) calculations show support to the
large-$N$ result and further identified the transition from VBS to
the antiferromagnetic N\'{e}el order as $N$ decreases
\cite{Harada2003,Kawashima2007,Paramekanti2007,Beach2009,Kaul2012}.
In literature, a phenomenological $t$-$J$-$\tilde{U}$ model
is often studied, in which $\tilde{U}$ is an effective onsite
repulsion but the antiferromagnetic super-exchange $J$ is put
by hand \cite{Yuan2005}.
At $\tilde{U}=0$, it becomes an unrestricted $t$-$J$ model which has
been studied by QMC on both the square lattice \cite{Assaad2005}
and the honeycomb lattice \cite{Lang2013}.
In the latter, the long-range antiferromagnetic N\'{e}el order
was found in the SU($2$) case, while the plaquette
VBS (pVBS) and the columnar VBS (cVBS) orders were identified
in the SU(4) and SU($2N\ge 6$) cases, respectively.

However, the SU($2N$) Heisenberg model completely neglects charge
fluctuations, while the unrestricted $t$-$J$ model misses the essential
Mott-physics originated from the onsite repulsion $U$.
The values of $J$ typically used in the unrestricted $t$-$J$ model
are overestimated compared to $J=4t^2/U$ arising
from the 2nd order perturbation super-exchange process based
on the Hubbard model.
Therefore, both the Heisenberg model and the unrestricted $t$-$J$
model cannot capture the rich competition and interplay between
charge and spin degrees of freedom in the entire interaction
region from the Dirac semi-metal phase to the Mott-insulating
phase.

In this paper, we study the original SU($2N$) Hubbard model on the honeycomb
lattice which correctly captures physics in both spin and charge channels.
Therefore, our results go beyond those based on the SU($2N$)
Heisenberg models and the unrestricted $t$-$J$ model.
Considering both the honeycomb optical lattice \cite{tarruell2012} and
the SU($2N$) ultra-cold fermions are already experimentally realized,
our simulations will provide helpful guidance for the future
experimental study for exotic quantum states.

We employ the projector determinant QMC method which is free of
the sign problem at half-filling \cite{Hirsch1985,White1989,Assaad2008}.
The non-perturbative nature of the QMC method is capable of describing
both charge fluctuations at intermediate interaction
region and Mott-physics in the strong interaction region.
As $U$ increases, quantum phase transitions from the Dirac semi-metal
phases to the Mott-insulating phases are identified for both SU(4)
and SU(6) cases.
The cVBS orders develop in the Mott insulating phases but both the
antiferromagnetic N\'{e}el and current-loop orders are absent.
The strength of the cVBS orders first grows and then drops in the
large-$U$ region due to the suppression of the overall kinetic
energy scale.
Meanwhile, both the single particle and spin gaps open in the cVBS phases.
All these results can be understood as a consequence of the competition
between the weak-$U$ itinerant physics and large-$U$ Mott physics.
Since the cVBS and pVBS share the same order parameter except
a $\pi$-phase difference, the transitions in general should
be 1st order based on the Ginzburg-Landau analysis.
Nevertheless, we also show that the transitions at zero temperature
may be continuous 2nd order due to the coupling
between the cVBS order and the gapless Dirac fermions.

The rest of this paper is organized as follows.
In Sect.\ref{sect:model}, the SU($2N$) Hubbard models are introduced,
and possible VBS ordering patterns are analyzed.
Throughout this article, we focus on two concrete examples with
SU($4$) and SU($6$) symmetries.
A new technique of distinguishing different VBS patterns is employed.
Other parameters for QMC simulations are also presented.
In Sect.\ref{sect:gap},  transitions from the semi-metal
phases to the Mott-insulating phases are studied by calculating
the dynamical properties of the system, including both the
single-particle gap and the spin gap.
In Sect.\ref{sect:order}, the competition between the N\'{e}el
ordering and the cVBS ordering is studied.
In Sect. \ref{sect:current}, the absence of the current
loop ordering is shown.
In Sect.\ref{sect:transition}, the nature of the phase transitions
between the gapless Dirac semi-metal phase and the gapped
cVBS phase is discussed.
Conclusions are made in Sect.\ref{sect:concl}.

\section{Model and Method}
\label{sect:model}

\subsection{SU($2N$) Hubbard model on honeycomb lattice}
\label{sect:model_order}
\begin{figure}[tp]
\includegraphics[width=\columnwidth]{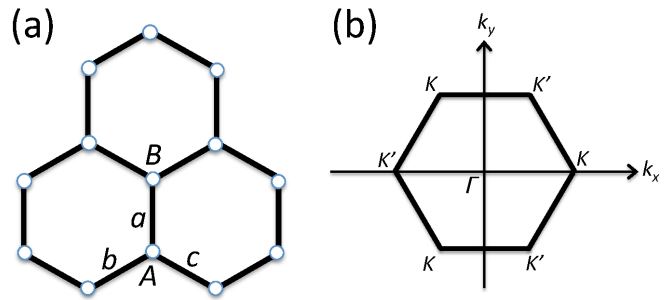}
\caption{$a$) The honeycomb lattice and the bond orientations.
($b$) The Brillouin zone. The Dirac points are located at
$\vec K$ and $\vec K^\prime$ at $(\pm \frac{4\pi}{3\sqrt 3 a_0},0)$.
The three $\vec K$($\vec K^\prime$) points are
equivalent to each other.
}
\label{fig:honeycomb}
\end{figure}

The honeycomb lattice is a bipartite but non-Bravais lattice.
Each unit cell consists two sites belonging to the $A$ and $B$-sublattices,
respectively.
Each site in the $A$-sublattice connects three bonds to its three
$B$-neighbors denoted as $a$, $b$ and $c$ as shown in Fig.
\ref{fig:honeycomb} ($a$).
The Brillouin zone can be represented as a regular hexagon
with the edge length $\frac{4\pi}{3\sqrt 3 a_0}$
as shown in Fig.~\ref{fig:honeycomb} ($b$),
where $a_0$ is the nearest-neighbor bond length.
It is well-known that the single-particle spectra on the honeycomb
lattice exhibit two gapless Dirac cones located at $\vec K$ and
$\vec K^\prime$, which is protected by the $D_{6}$ point group symmetry.

We employ the following SU($2N$) Hubbard model defined in the
honeycomb lattice as
\bea
H&=&-t\sum_{i \in A, \hat e_j; \alpha}
\left(c_{i\alpha}^{\dagger}c_{i+\hat e_j, \alpha}+h.c.\right)
\nn \\
&+&\frac{U}{2}\sum_{i\in A \oplus B}\left(n_{i}-N\right)^{2},
\nn \\
\label{eq:hubbard}
\eea
where $\hat e_j$'s with $j=a,b,c$ represent
unit vectors along three bond orientations, respectively;
$\alpha$ is the spin index taking values from 1 to $2N$;
$n_{i}=\sum_{\alpha}c_{i\alpha}^{\dagger}c_{i\alpha}$ is the total particle
number operator on site $i$;
$t$ is the hopping integral which is
set as energy unit throughout the paper; $U$ is the onsite Coulomb repulsion.
Eq. \ref{eq:hubbard} has already been set at the particle-hole
symmetric point such that $\avg{n_i}=N$, thus the chemical potential
does not appear explicitly.
The convention of $U$ is defined as follows:
In the atomic limit of $t/U\rightarrow 0$, on the half-filled background,
if a single particle is moved from one site to another,
the excitation energy is $U$, independent of $N$.

The SU($2N$) generators on each site $i$ are defined as
\bea
S_{\alpha\beta}(i)=c_{i,\alpha}^{\dagger}c_{i,\beta}-
\frac{\delta_{\alpha\beta}}{2N}\sum_{\gamma=1}^{2N}c_{i,\gamma}^{\dagger}c_{i,\gamma},
\label{eq:sun_generator}
\eea
which satisfies the constraint of $\sum_{\alpha} S_{\alpha\alpha}(i)=0$.
Compared with the representation in terms of the Gellman matrix,
the generators $S_{\alpha\beta}(i)$ satisfy the commutation relation in a
simpler form as
\bea
[S_{\alpha\beta}(i), S_{\gamma\delta}(j)]
=\delta_{ij}\left[\delta_{\beta\gamma}S_{\alpha\delta}(i)
-\delta_{\alpha\delta}S_{\gamma\beta}(i)\right].
\eea
The two-point equal-time SU($2N$) spin-spin correlation function is
defined as
\bea
S_{spin}(i,j)=\sum_{\alpha,\beta}\langle
S_{\alpha\beta}(i) S_{\beta\alpha}(j)\rangle,
\label{eq:SS}
\eea
where $\avg{}$ means the expectation value evaluated over the ground state.
The staggered spin structure factor is defined as
\bea
S_{stag}(L)=\frac{1}{2L^2}\sum_{ij}(-1)^{i+j}S_{spin}(i,j),
\label{eq:neel_cor}
\eea
where $L$ is the linear system size and $2L^2$ is the number of lattice
sites.
The long-range N\'eel order is given by
\bea
M_{nl}= \lim_{L\rightarrow\infty} \sqrt{\frac{1}{2L^{2}}S_{stag}(L)} .
\label{eq:neel_moment}
\eea

\begin{figure}[t]
\includegraphics[width=\columnwidth]{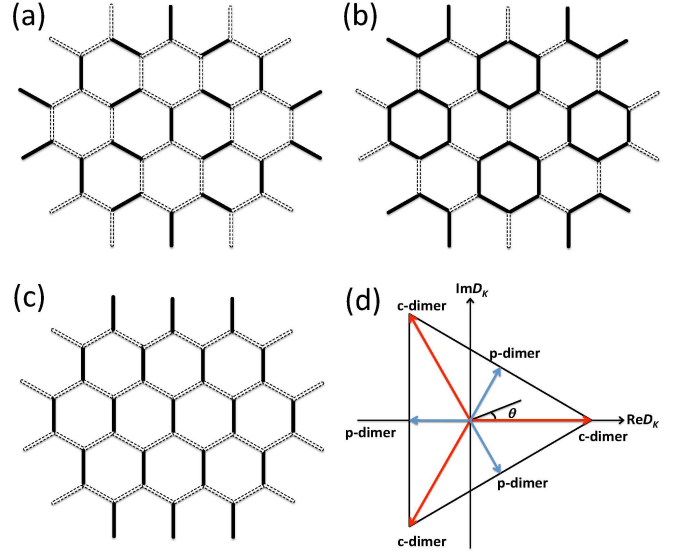}
\caption{Possible VBS configurations: ($a$) cVBS and ($b$) pVBS
break translational symmetry and exhibit a $\sqrt 3\times \sqrt 3$
super-unit cell;
($c$) the sVBS configuration maintains translational symmetry.
($d$) The argument $\theta$ of the VBS order parameter $D_{K,1}$.
The values of $\theta=0,\frac{2}{3}\pi, \frac{4}{3}\pi$ represent
the ideal cVBS order depicted in ($a$), and the values of
$\theta=\pi, \frac{5}{3}\pi, \frac{1}{3}\pi$ represent
the ideal pVBS order depicted in ($b$).
}
\label{fig:dimer_pat}
\end{figure}

Next we analyze the possible VBS ordering pattern on
the honeycomb lattice.
Typically, there are three non-equivalent VBS configurations:
columnar VBS (cVBS), plaquette VBS (pVBS), and the staggered
VBS (sVBS) as depicted in Fig. \ref{fig:dimer_pat}
($a$), ($b$), and ($c$), respectively.
The $s$-dimer ordering still maintains the two-site unit cell.
Among them, sVBS does not break translational symmetry but breaks
the six-fold rotational symmetry down to two-fold.
It shifts the locations of Dirac points away from $\vec K$
and $\vec K^\prime$.
The sVBS does not open gap unless its magnitude
is comparable to $t$, thus it is not favored.
The $c$- and $p$VBS orders exhibit the $\sqrt 3\times \sqrt 3$ structure,
and their symmetry patterns are the same.
They enlarge the unit cell to six sites but still maintain the six-fold
rotational symmetry.
Both the $c$VBS and $p$VBS orderings are at the wave vector
$\vec K$ or $\vec K^\prime$.
Infinitesimal $c$VBS and $p$VBS orders open the single particle gap
and render the semi-metal with Dirac cones to an insulator.
Below we will only consider the cVBS and pVBS orderings.

To distinguish different VBS dimer orderings, without loss of generality,
we consider each site $i$ of the $A$-sublattice and define
its three nearest neighboring bonds $d_{i,\hat{e_j}}$ as
\bea
d_{i,\hat{e}_j}=\frac{1}{2N}\sum_{\alpha=1}^{2N}\left(c_{i\alpha}^\dagger
c_{i+\hat{e}_j,\alpha}+h.c.\right),
\eea
where $j$ represents the three bonds shown in Fig. \ref{fig:honeycomb}
($a$).
Following Ref. \onlinecite{pujari2015},
we define the complex order parameters $D_{K,m}$ with
$m=0,\pm 1$,
\bea
D_{K,m}(L)=\frac{1}{L^{2}}
\sum_{i\in A} \left(d_{i,\hat e_a} +\omega^m d_{i,\hat e_b}
+\omega^{2m} d_{i,\hat e_c}\right)e^{i\vec K\cdot \vec r_{i}},
\nn \\
\label{eq:pc_dim}
\eea
in which $\omega=e^{i\frac{2}{3}\pi}$, and $m=0,\pm 1$.
If we change $\vec K$ to $\vec K^\prime$ in Eq. \ref{eq:pc_dim},
it does not give rise to new orders since
$D_{K,m}=D_{K^\prime,-m}^*$.
The cases with $m=0, -1$ correspond to the trimer orderings.
$D_{K,m=1}$ corresponds to the $p$ and $c$-dimer orderings.
The difference between $m=\pm 1$ is because the phase factors
in Eq. \ref{eq:pc_dim} are associated with $\vec K$ which
already breaks the equivalence between the chiral indices
$m=\pm 1$.

For simplicity, we define the following dimer strength parameter
which only keeps correlations among bonds along same directions as
\bea
\mbox{dim}_K &=&
\lim_{L\rightarrow +\infty}
\frac{1}{L^{2}}\sqrt{\sum_{i,i^\prime\in A; \hat e_j}
e^{iK\cdot(r_{i}-r_{i^\prime})}\langle d_{i,\hat e_j}d_{i^\prime,\hat e_j}\rangle} \nn \\
&=& \lim_{L\rightarrow+\infty}
\frac{1}{\sqrt{3}L^{2}} \sqrt{\sum_{m=0,\pm1} D^*_{K,m}(L) D_{K,m}(L)}.
\label{eq:dimK}
\eea
As shown in Appendix \ref{app:dimer},
the cases of $m=0$ and $-1$ do not exhibit long-range ordering
at half-filling.
Then after the finite size scaling, dim$_K$ actually represents
the cVBS and pVBS orderings as
\bea
\mbox{dim}_K &=&
\lim_{L\rightarrow+\infty}\frac{1}{\sqrt{3} L^{2}} \sqrt{D^*_{K,1}(L) D_{K,1}(L)}.
\eea

Although the cVBS and pVBS states cannot be distinguished from
the magnitudes of their common order parameter,
the argument of the complex order parameter $D_{K,1}$ exhibits different
patterns as shown in Fig. \ref{fig:dimer_pat} ($d$).
For example, for the ideal cVBS states,
$\arg(D_{K,1})=0,\frac{2}{3}\pi$,
or, $\frac{4}{3}\pi$, which are related to each other by translational
or rotational symmetries.
In contrast, for the ideal pVBS states,
$\arg(D_{K,1})=\pi,\frac{5}{3}\pi$,
or,  $\frac{1}{3}\pi$, which distribute inversion
symmetrically with respect to those of the cVBS states.
To see the difference between cVBS and pVBS more clearly,
we define the following parameter
\bea
W=\int dz dz^* P(z,z^*) \cos 3\theta ,
\label{eq:histW}
\eea
where $z=D_{K,1}$, $\theta=\arg(z)$, and $P(z,z^*)$ is the probability
density appearing in the Monte-Carlo sampling satisfying
$\int dz dz^* P(z,z^*)=1$.
For the ideal cVBS state, $W=1$, while for the ideal pVBS state, $W=-1$.

\subsection{Parameters of QMC simulations}
We apply the zero temperature projector QMC (PQMC) method
\cite{Assaad1998,Assaad2008} in the determinant formalism
\cite{Hirsch1985}.
To preserve translational symmetry, the honeycomb lattice is subject
to periodic boundary condition in real space.
The half-filled SU($2N$) Hubbard model is free of the minus-sign problem in
bipartite lattices as shown in Appendix \ref{append:sign}.

The trial wave function is chosen as the ground state of the non-interacting
part in Eq.~\ref{eq:hubbard} with a small flux added to break the degeneracy at the Dirac points~\cite{Meng2010}.
The scalings of $\Delta\tau\to0$ and $\beta\to\infty$ scaling of
physical observables are performed in appendix \ref{app:tau}
and appendix \ref{app:beta}, respectively.
We find simulation parameters $\Delta\tau=0.05$ and $\beta=40$ for QMC are
sufficient to give accurate ground state properties of the system.
The measurements of physical observables are performed close to $\beta/2$
after the projection to the ground state.
For a typical largest $L=15$ system, we run 500 warm-up QMC steps, followed
by at least 20 QMC bins with 700 measurements inside each bin.
And for relatively smaller size, more warm-up steps and measurement steps
are used.

\section{Quantum phase transition between massless Dirac semi-metal and
cVBS insulating phase}
\label{sect:gap}

In this section, we study the quantum phase transitions from
the semi-metal phase of Dirac fermions to the gapped insulating phases
both for the SU(4) and SU(6) cases.
To identify the transition, we focus on the single-particle gap $\Delta_{sp}$
and the spin gap $\Delta_{\sigma}$ as a function of interaction $U$.

Physics of the honeycomb lattice is very different from that in the square
lattice, in both the weak and strong coupling limits.
In the weak coupling regime, at half-filling, the honeycomb lattice
possesses robust massless Dirac fermion spectrum and the interaction
effects are weakened by the vanishing of density of states.
On the other hand, the square lattice has Fermi surface nesting and
van Hove singularities where the density of states diverges logarithmically.
Hence, the square lattice system is unstable at infinitesimally small
interaction and the single-particle gap $\Delta_{sp}$ opens from
$U=0^{+}$.
The corresponding antiferromagnetic insulator is a typical Slater-type
insulator with a spin-density-wave order\cite{Hirsch1985}.
In the strong coupling regime at $U\to \infty$, the Mott physics gives
rise to a large gap as $\Delta_{sp}\approx \frac{U}{2}$
\cite{Hubbard1963,*Hubbard1964}.
The honeycomb lattice has the smallest coordination number $z=3$ among
all of the 2D lattices.
Compared to the square lattice with $z=4$, the charge excitations
in the honeycomb lattice are more difficult to delocalize,
and thus its Mott physics is more robust.
In comparison, both the weak and strong coupling regimes
in the honeycomb lattice are more extended in terms of $U/t$
than in the square lattice.
Hence, the intermediate coupling regime in the honeycomb
lattice should be greatly suppressed.

\subsection{Single-particle gap $\Delta_{sp}$}
\label{sec:spgap}
We measure the single-particle gap $\Delta_{sp}$ in QMC to monitor the
transition from the semi-metal to Mott insulator.
Before presenting the QMC results, let us discuss an intuitive picture
to obtain the critical $U_c$ for the opening of the single
particle gap \cite{Zhou2014}.
Suppose that we start with the strong coupling limit and add one more
fermion on a site.
After turning on the hopping $t$, the extra charge hops to its nearest
neighbors.
The number of possible hopping processes are $zN$,
thus the band width $W_b\approx 2zNt$.
The single-particle gap decreases approximately as
\bea
\Delta_{sp}\approx\frac{U}{2}-\frac{W_b}{2}\approx \frac{U}{2}- z N t,
\label{eq:bdwdth}
\eea
leading to $U_c/t\approx 2zN$.
This argument yields the trend that the larger the fermion flavors $2N$,
the stronger critical $U_c$ is needed to open the $\Delta_{sp}$,
in an approximate linear way.
This trend is qualitatively consistent with our QMC results.

\begin{figure}[htb]
\includegraphics[width=\linewidth]{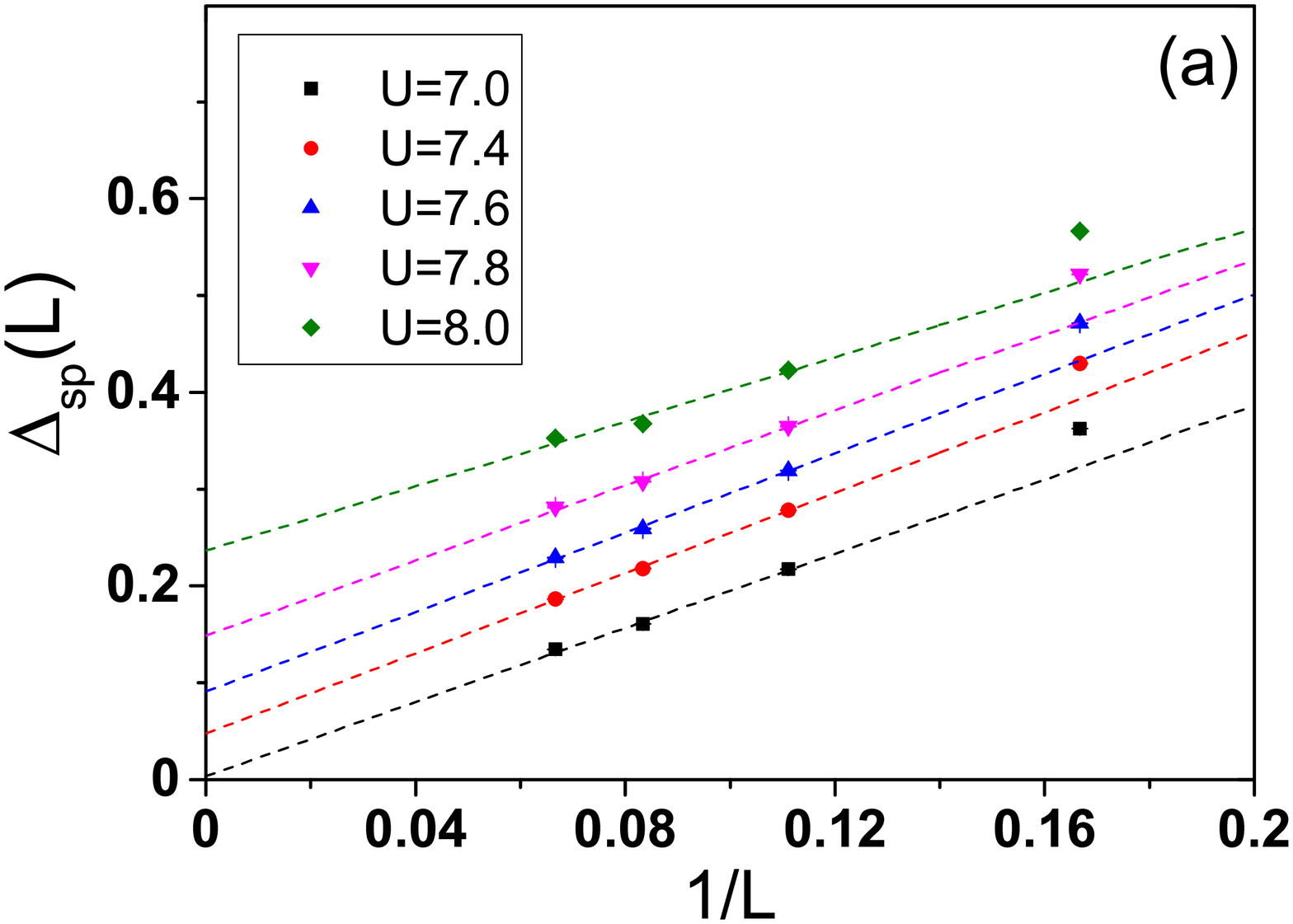}
\includegraphics[width=\linewidth]{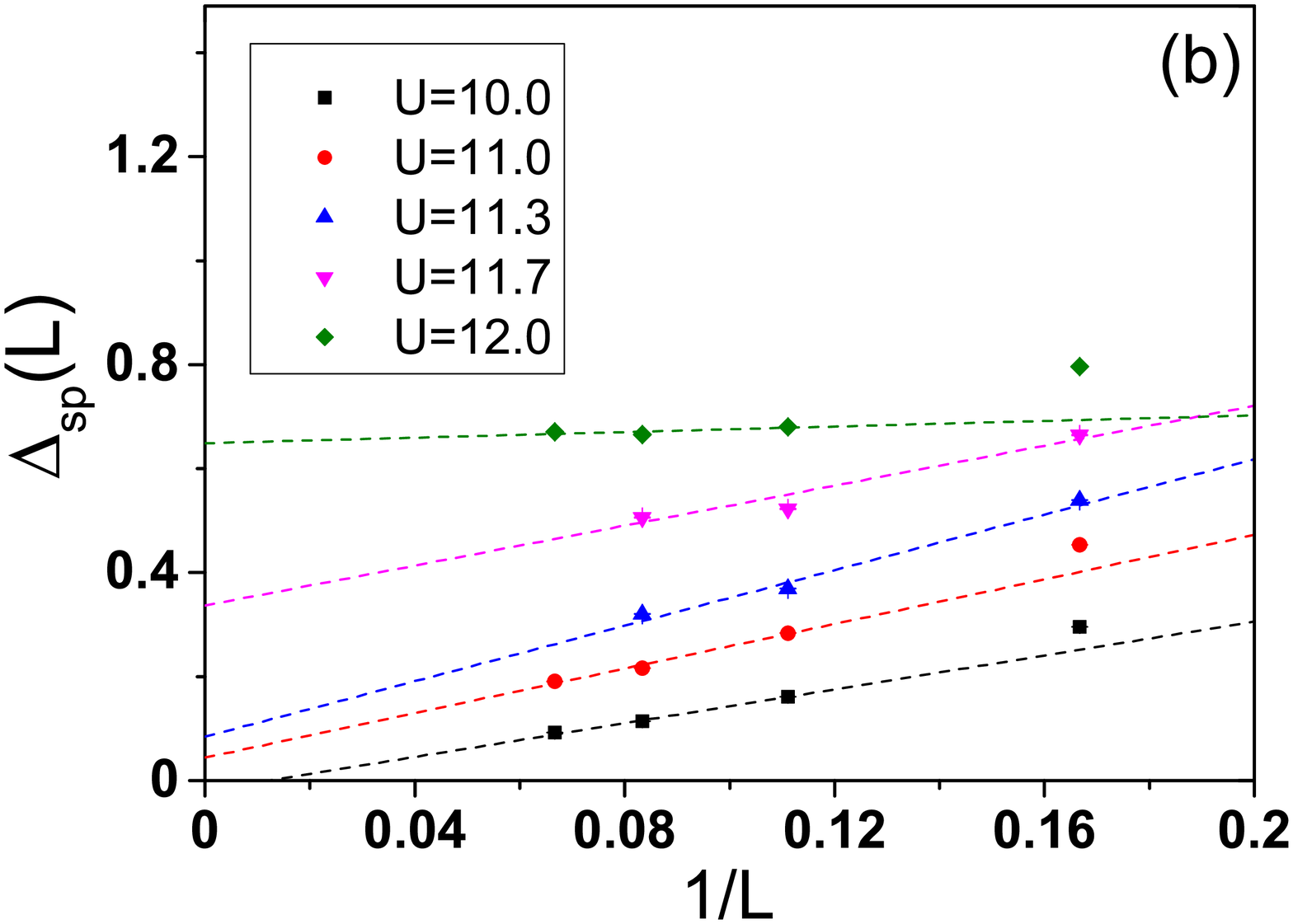}
\caption{Finite size scalings of the single-particle gap
$\Delta_{sp}$ for ($a$) the SU($4$)  and ($b$) SU($6$)
Hubbard models with different values of $U$.
The linear fitting is applied using the last three data points
for each $U$.
Error bars of the finite size $\Delta_{sp}$ are smaller than the symbols.
}
\label{fig:sun_gap}
\end{figure}

Since the minimal single-particle gap is located at the Dirac points
$\vec{K}$ and $\vec{K'}$, we consider the imaginary-time displaced
Green's function at $\vec K$ as
\bea
G(\vec K,\tau)=\frac{1}{2L^{2}}\sum_{i,j\in A\oplus B}
G(i,j,\tau)e^{i\vec K\cdot (\vec r_{i}-\vec r_{j})},
\eea
where
\bea
G(i,j,\tau)=\sum_{\alpha}\langle\Psi_{G}|c_{i,\alpha}(\tau)
c_{j,\alpha}^{\dagger}(0)|\Psi_{G}\rangle,
\eea
and $|\Psi_{G}\rangle$ is the ground state.
After a long imaginary-time displacement, the asymptotic behavior of
$G(\vec K,\tau)$ scales as $e^{-\tau\Delta_{sp}}$, and $\Delta_{sp}$ can
then be obtained through fitting the slope of the
$\ln G(\vec K,\tau)$.
In Appendix \ref{append:single_gap}, the raw data of $\ln G(\vec K,\tau)$
are presented and the values of $\Delta_{sp}$ are extracted for different $L$.
The finite size scalings of $\Delta_{sp}(L)$ in the SU($4$) case for
different $U$ are presented in Fig. \ref{fig:sun_gap} $(a)$.
The critical interaction for the opening of $\Delta_{sp}$ is
$U_{c}(2N=4)\approx 7$.
For the SU($6$) case, the same finite size scaling is presented
in Fig. \ref{fig:sun_gap} $(b)$, and $\Delta_{sp}$ becomes finite
around $U_c (2N=6)\approx 10$.
In contrast, in the SU(2) case, the critical value of $U_{c}(2N=2)\approx 3.7$
in literature  \cite{Meng2010,Assaad2013}.

\begin{figure}[htb]
\includegraphics[width=\linewidth]{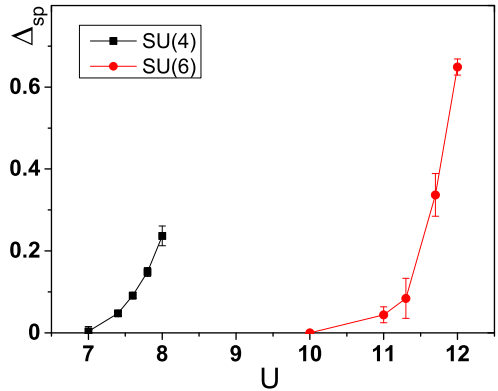}
\caption{The single particle gap $\Delta_{sp}$ {\it v.s.} $U$ for
the half-filled SU($4$) and SU($6$) Hubbard model in the honeycomb lattice.
}
\label{fig:single_gap}
\end{figure}

The above QMC results show that with increasing $2N$, $U_c$ also increases,
exhibiting an approximate linear relation in agreement with the result
based on Eq. \ref{eq:bdwdth}.
Nevertheless, Eq. \ref{eq:bdwdth} significantly overestimates the
values of $U_c$ compared with the accurate QMC results.
The reason is that the charge excitations on the Mott insulating
background are significantly incoherent, which suppresses the band
width $W_d$ in Eq. \ref{eq:bdwdth}.
This intrinsically many-body effect is beyond the single-particle
picture as captured by the QMC simulations.

In Fig. \ref{fig:single_gap}, the relations $\Delta_{sp}$ {\it v.s}.
$U$ are plotted for both the SU($4$) and SU($6$) cases.
It is more difficult to open the single-particle gap in SU($6$) than SU($4$).
Nevertheless, $\Delta_{sp}$ increases more quickly in the SU($6$) case
after passing $U_c$.
As will be seen in Sec. \ref{sect:order}, the VBS order parameter
$\mbox{dim}_K$ is much stronger in the SU($6$) case than in the SU($4$) case.
The enhancement of the VBS ordering suppresses the mobility of charge
excitations, i.e., the band width $W$ in Eq. \ref{eq:bdwdth}.
As a result, $\Delta_{sp}$ grows more quickly when $U$ just passes $U_c$
in the SU(6) case.

\subsection{Spin gap $\Delta_\sigma$}
\label{sect:spin_gap}

In order to further understand the nature of the gapped phases at $U>U_c$,
spin gaps $\Delta_\sigma$ are calculated in both SU($4$) and SU($6$) cases.
Before presenting QMC data, we can solve a two-site problem to
understand the overall behavior of $\Delta_\sigma$ as varying $U$ and
$N$ in the strong coupling regime.
In this regime, the magnetic properties can be effectively described by
the SU($2N$) Heisenberg model as
\bea
H_{ij}^J=\frac{J}{2}\sum_{\alpha\beta} \Big\{
S_{\alpha\beta}(i) S_{\beta\alpha}(j)-\frac{1}{2N}n_i n_j\Big\}.
\label{eq:superexchange}
\eea
Since each site is half-filled, each SU($2N$) spin operator
lies in the self-conjugate representation.
The superexchange energy scales as $J=4t^2/U$ at the 2nd order
perturbation theory.
After simple calculations, the ground state of this two-site problem
is an SU($2N$) singlet with $E_0=-\frac{N(N+1)}{2}J$,
and the first excited state is in the SU($2N$) adjoint representation
which is an analogue of the triplet excitation in the SU(2) case.
The spin gap of this two-site problem can be calculated as
\bea
\Delta_\sigma \approx 4Nt^2/U,
\label{eq:bd_spgap}
\eea
which shows that $\Delta_\sigma$ is enhanced by increasing the fermion component
number $N$, but suppressed by increasing $U$.
Although Eq. \ref{eq:bd_spgap} is only accurate for a single
bond of two sites,
it still can qualitatively yield an intuitive picture for $\Delta_\sigma$
in systems with VBS orderings in both SU($4$) and SU($6$) cases.

To extract the spin gap $\Delta_\sigma$, we employ the imaginary-time
displaced SU($2N$) spin-spin correlation function defined as
\bea
G_{\sigma}(\tau)=\frac{1}{2L^{2}}\sum_{i,j,\alpha,\beta}
\Big\{\left(-1\right)^{i+j}\langle
S_{\alpha\beta,i}(\tau)S_{\beta\alpha,j}(0)\rangle\Big
\}. \ \ \
\eea
Similar to the case of the single-particle gap, the value of the spin
gap $\Delta_{\sigma}$ can be obtained by fitting $\ln G_{\sigma}(\tau)$.
In Appendix \ref{append:spingap}, the raw data of $\ln G_\sigma(\tau)$
are plotted and spin gaps are extracted.

\begin{figure}[htb]
\includegraphics[width=\columnwidth]{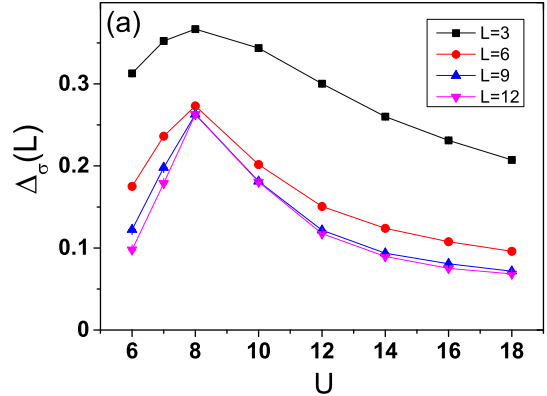}
\includegraphics[width=\columnwidth]{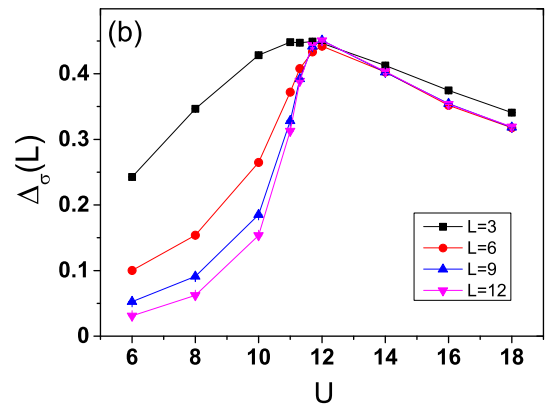}
\caption{Spin gap $\Delta_\sigma$ for finite size systems of the SU($2N$) Hubbard model in the honeycomb lattice: ($a$) the SU(4) case and
($b$) the SU(6) case. Error bars are smaller than the symbols.}
\label{fig:spingapL}
\end{figure}

The spin gap $\Delta_\sigma$ v.s. $U$ at different system sizes $L$
are presented in Fig.\ref{fig:spingapL} ($a$) and ($b$) for the
SU($4$) and SU($6$) cases, respectively.
At $U<U_c$, the systems are in the semi-metal phase, and thus the
spin gaps vanish in the thermodynamic limit $L\rightarrow\infty$.
In both the SU(4) and SU(6) cases, the relations of $\Delta_\sigma$
at finite size $L$ v.s. $U$ are non-monotonic:
They reach maxima at values of $U$ slightly larger than $U_c$, and after
that, they decrease as $U$ further increases, in agreement with
intuitive understanding in Eq. \ref{eq:bd_spgap}.
The positions of the peak in $\Delta_{\sigma}$ do not shift with $L$, and
the $\Delta_{\sigma}$ converges quickly with $L$ after
passing the peak values, so the Fig.~\ref{fig:spingapL}
($a$) and ($b$) show that spin gaps are non-zero in thermodynamic
limit for the gapped phase and there is no N\'{e}el order at $U>U_c$
(also see Sec.~\ref{sect:neel} below).
The opening of $\Delta_\sigma$ at $U>U_c$ can be attributed to the
developing of the VBS ordering as will be shown in Sect. \ref{sect:dimer}.

The behavior of the spin gap $\Delta_{\sigma}$ is also very different
from the single-particle gap $\Delta_{sp}$.
The latter involves charge excitations while the former does not,
thus the single-particle gap grows with $U$ monotonically and approaches
$\frac{U}{2}$ in the strong coupling limit.
As analyzed in Sec.~\ref{sec:spgap}, the single-particle gap $\Delta_{sp}$
is weakened by increasing the fermion component number $2N$.
In contrast, the spin gap is enhanced as shown in Eq.~\ref{eq:bd_spgap}
in the strong coupling regime.
Although in the strong coupling regime $\Delta_{sp}\gg \Delta_{\sigma}$,
$\Delta_{sp}$ is comparable with $\Delta_{\sigma}$
in the intermediate coupling regime.
This means that the spin and charge fluctuations are intertwined
with each other, and there is no clear separation between them
for low energy physics~\cite{Meng2010}.

\section{N\'{e}el v.s. VBS as the ground state}
\label{sect:order}

In this section we address the possible competition between the
antiferromagnetic N\'{e}el order and the VBS dimer order, and explain
why there is no N\'eel order as the ground state in the SU($4$)
and SU($6$) cases at $U>U_c$.

\begin{figure}[tp]
\includegraphics[width=\columnwidth]{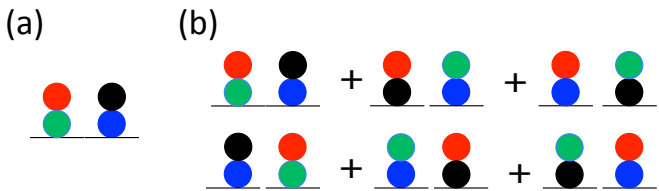}
\caption{($a$) The classic N\'{e}el and ($b$) quantum dimer configurations
across a bond for the case of $2N=4$.
In ($a$), fermion components on one site are $1$ to $N$,
and on the other site are $N+1$ to $2N$.
In ($b$), two sites form an SU($2N$) bond singlet.
The exchange energy gain per bond in the N\'{e}el case is $2N t^2/U$, and
that for a singlet bond is $2N (N+1) t^2/U$.
}
\label{fig:dimer_neel}
\end{figure}

Again, let us first discuss an intuitive picture to explain why increasing
the fermion components favors the VBS order and suppresses the N\'{e}el
order.
We consider a two-site problem of a single bond in the strong coupling
limit in which charge fluctuations can be neglected, and compare the
energy gains for the N\'{e}el and SU($2N$) bond singlet configurations
as depicted in Fig. \ref{fig:dimer_neel}.
To maintain the N\'{e}el configuration,  spin-flip process is not allowed.
One fermion of a certain species on the left site can  hop to the right,
and then it must hop back.
The total number of the virtual hopping processes is $2N$, and thus
the energy gain at the 2nd order perturbation level is $2N\frac{t^2}{U}$.
In contrast, if two sites form a SU($2N$) singlet, spin-flip processes
are allowed and the energy gain is $2N(N+1) \frac{t^2}{U}$.
Hence from the energy perspective, the SU($2N$) singlet is the ground
state over the N\'{e}el order, in the two-site problem.
Of course, on the 2D lattice, situation becomes more complicated,
the N\'{e}el state enjoys the advantage that every site can fit all
of its neighbors, while for VBS one site can only participate in
the formation of one singlet.
But, roughly speaking, as $2N$ increases much larger than the coordination
number $z$, we expect the VBS dimer order should win due to the enhanced
quantum spin fluctuations \cite{Lang2013,Wang2014}.
On the contrary, the N\'{e}el order is expected to win at small $N$,
such as in the SU($2$) case~\cite{Meng2010,Sorella2012,Assaad2013}.

\subsection{The N\'{e}el order}
\label{sect:neel}

\begin{figure}[htb]
\includegraphics[width=\columnwidth]{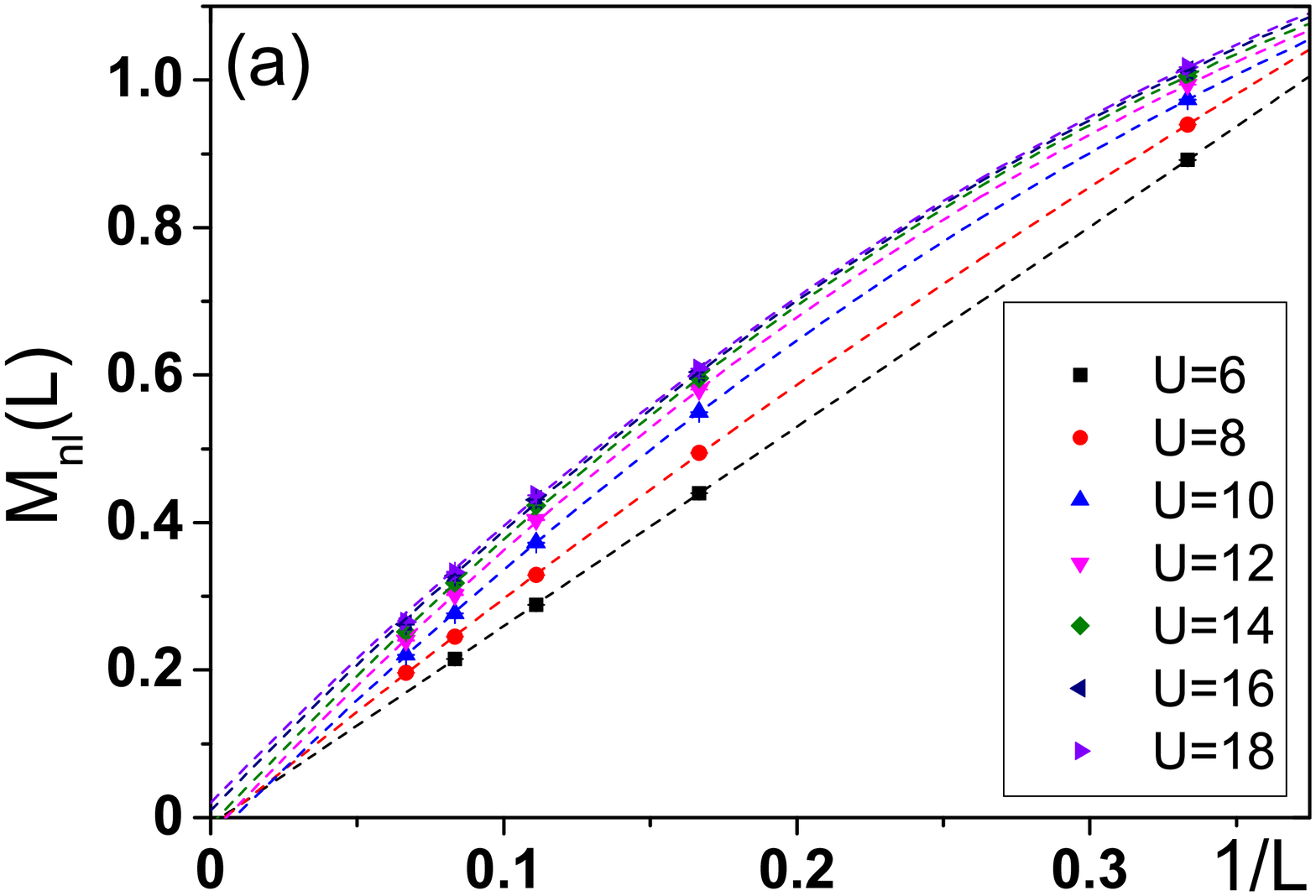}
\includegraphics[width=\columnwidth]{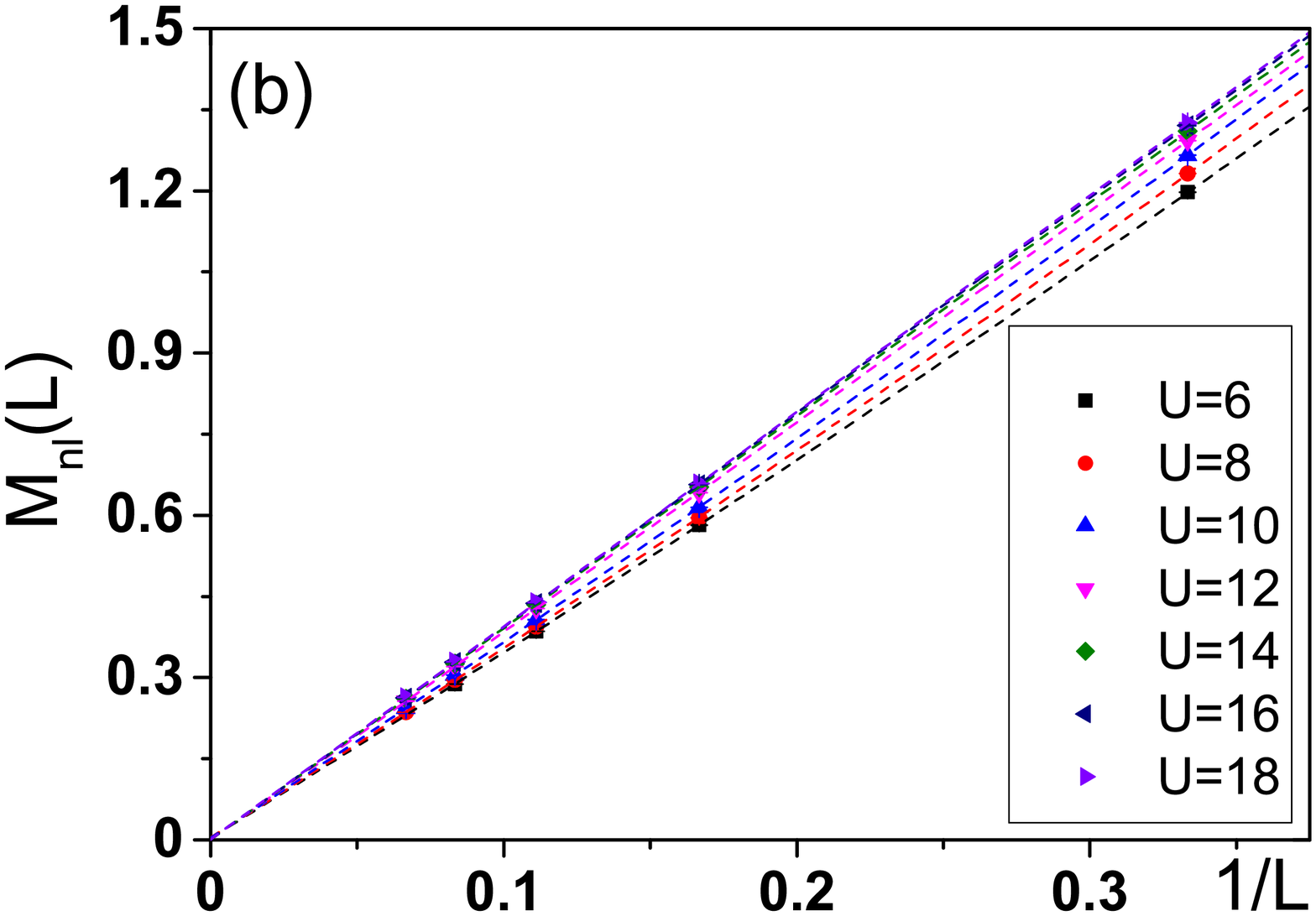}
\caption{The finite size scalings of the N\'{e}el order parameter
$M_{nl}$ against $1/L$ at different values of $U$: ($a$) the SU(4) case
and ($b$) the SU(6) case.
The quadratic polynomial fitting is used for $L$ from 3 to 15.
Error bars  are smaller than symbols.}
\label{fig:sun_neel}
\end{figure}

Previous QMC results for the SU(2) case \cite{Meng2010,Sorella2012,Assaad2013}
showed that the N\'{e}el order is the ground state at large $U$.
Our results of the SU(4) and SU(6) cases are quite different:
the long-range N\'{e}el order is absent.
The finite size scalings of the AF order parameter $M_{nl}$ defined
in Eq.~\ref{eq:neel_moment} for the SU(4) and SU(6) cases are presented
in Fig. \ref{fig:sun_neel} ($a$) and ($b$), respectively.
For the SU(4) case, the finite size scaling on $M_{nl}(L)$ shows
clear evidence of the vanishing of long-range N\'{e}el order at
intermediate values from $U=6$ to $14$.
The AF correlation is enhanced by further increasing $U$.
At $U=16, 18$, the finite-size scalings show very small residual
values as $L\to \infty$.
Nevertheless, it might well be an artifact of limited system sizes
($L$ is up to 15).
In particular, the curvatures of these $M_{nl}(L)$ curves are negative
and thus it is conceivable that they will finally converge to zero
as $L \to \infty$.
And also in appendix \ref{app:pin}, we find there is no N\'{e}el order for large $U$ cases by using pinning field method~\cite{Assaad2013}.
The situation in the SU(6) case is clearer: All the curves of $M_{nl}(L)$
exhibit a clear linear scaling in terms of $1/L$. We can safely conclude the absence of the long-range N\'{e}el order for all the values of $U$ from 6 to 18.

In both SU(4) and SU(6) cases, long-range N\'{e}el order does not
appear after the opening of the single-particle gap, which is
in agreement with the developing of non-zero spin gaps.
Compared to the N\'{e}el ordering in the SU(2) case in the honeycomb lattice,
the larger values of $2N=4,6$ enhance quantum spin fluctuations.
Let us compare with the result of SU($2N$) Hubbard model in the square
lattice \cite{Wang2014}.
The N\'{e}el order remains robust in the SU(4) case for the entire
parameter range of $0<U<20$ simulated.
Although the N\'{e}el order is weakened at large values of $U$, it
remains an open question whether it can persist to the limit of
$U\to \infty$.
As for the SU(6) case in the square lattice, the N\'{e}el order still
appears in the weak and intermediate coupling regimes ($0<U<15$),
and is replaced by the VBS dimer order in the strong
coupling regime.

\subsection{The VBS order}
\label{sect:dimer}

\begin{figure}[htb]
\includegraphics[width=\columnwidth]{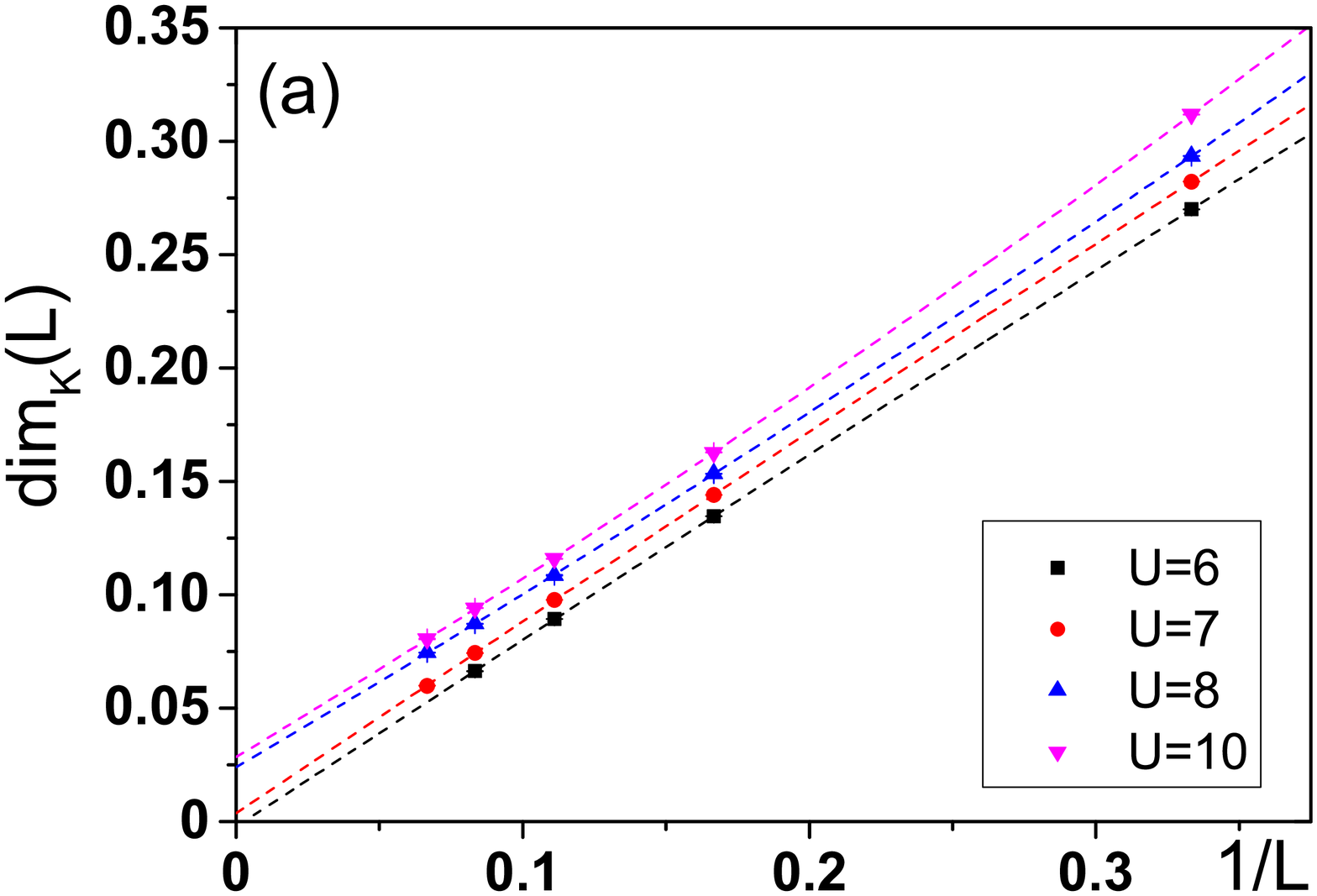}
\includegraphics[width=\columnwidth]{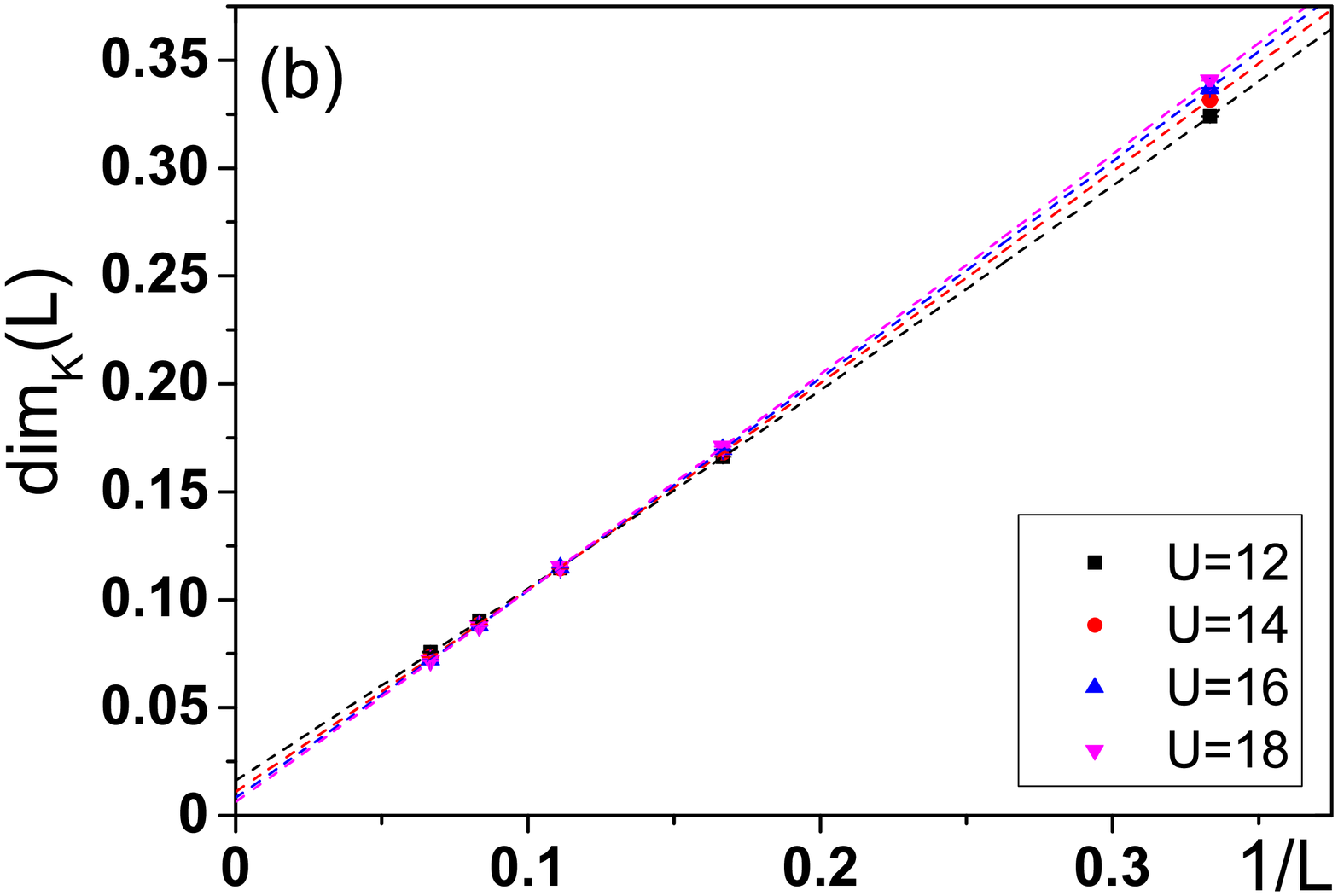}
\caption{Finite size scalings of $\mbox{dim}_{K}(L)$ {\it v.s.}
$1/L$ for the SU(4) Hubbard model: ($a$) $U=6,7,8,10$, and
($b$) $U=12, 14, 16, 18$.
The quadratic polynomial fitting is used, and error bars are
smaller than symbols.
}
\label{fig:su4_dimer}
\end{figure}

\begin{figure}[htb]
\includegraphics[width=\columnwidth]{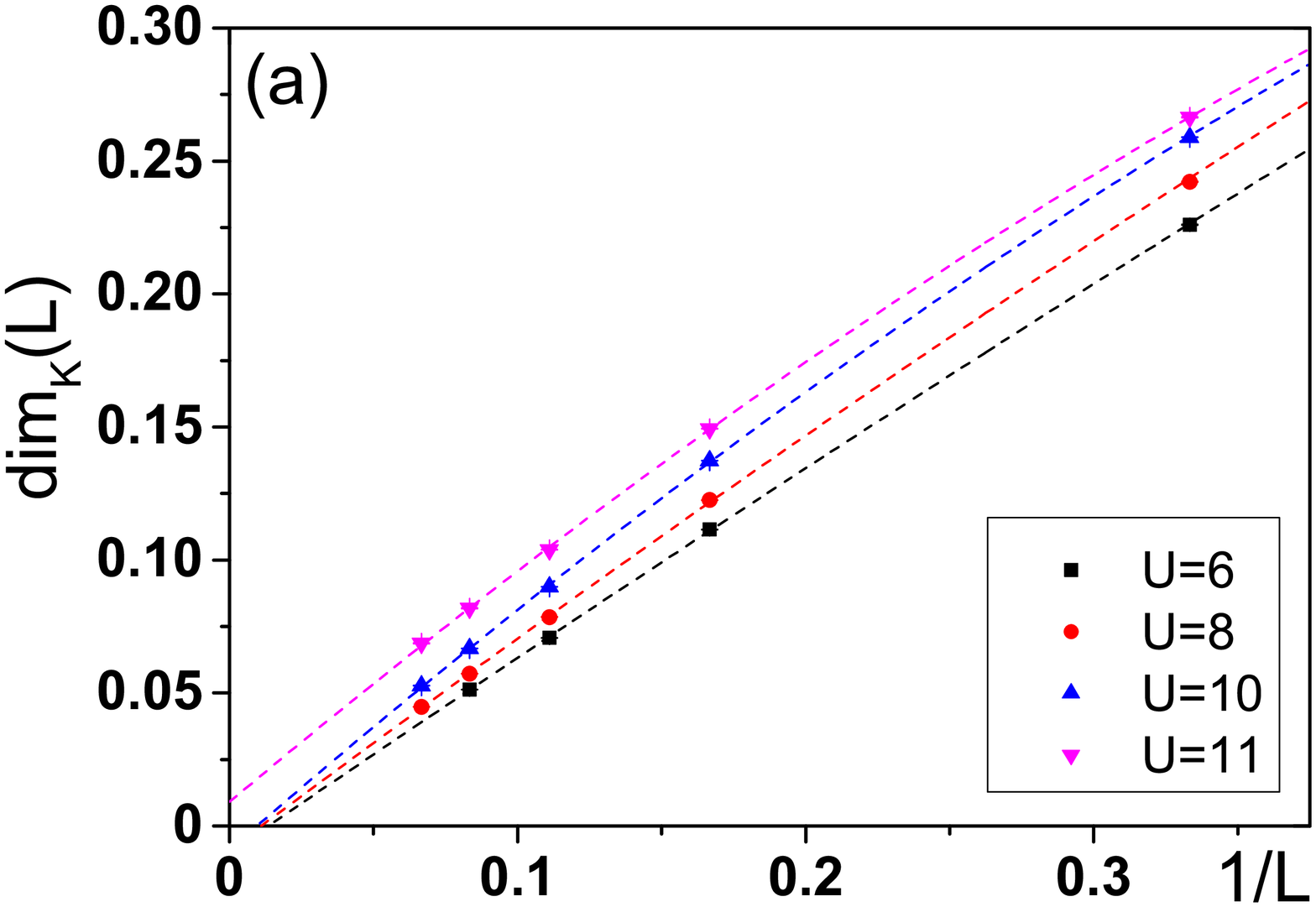}
\includegraphics[width=\columnwidth]{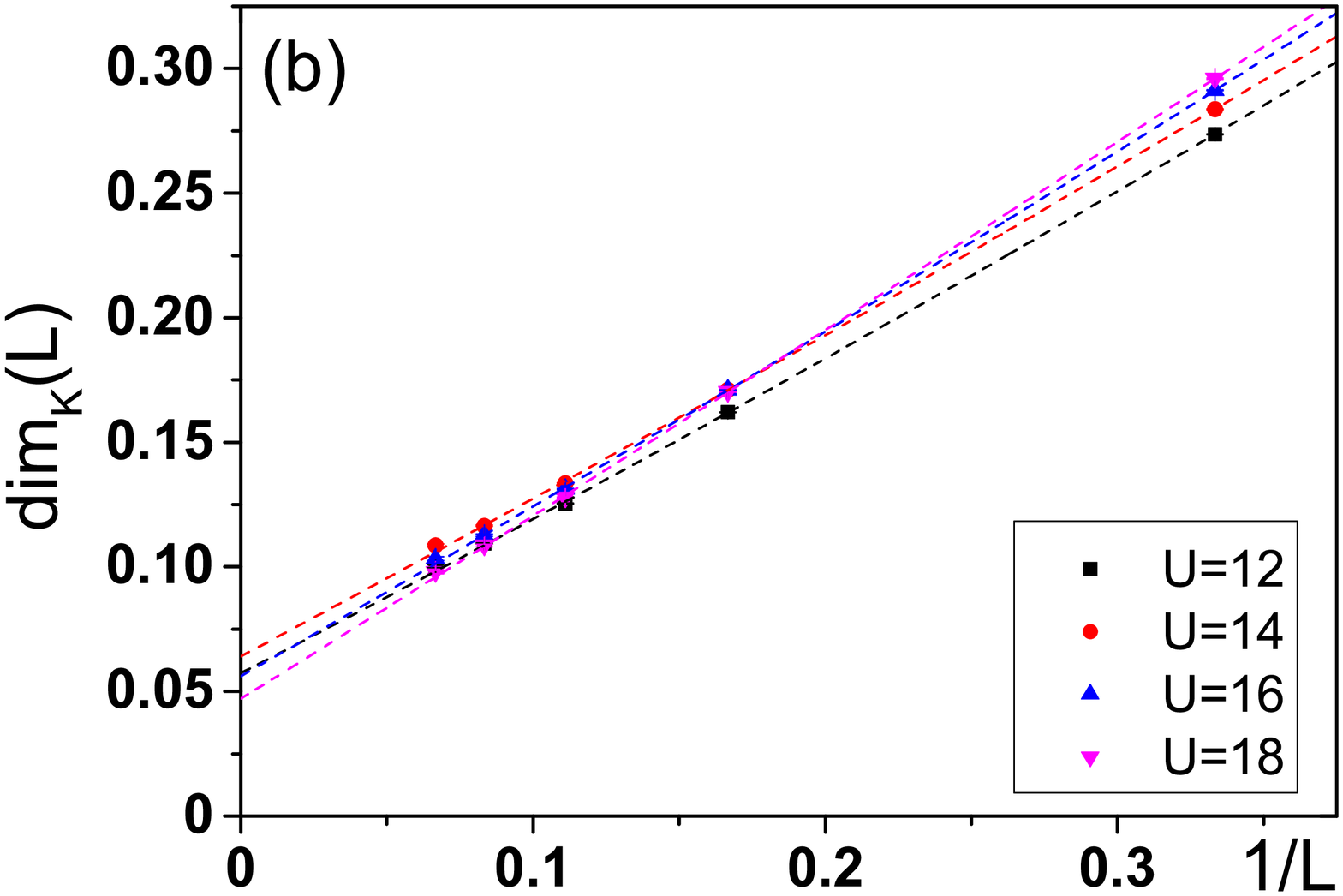}
\caption{Finite size scalings of $\mbox{dim}_{K}(L)$ {\it v.s.} $1/L$
for the SU(6) Hubbard model: ($a$) $U=6,8,10, 11$, and ($b$) $U=12, 14,
16, 18$.
The quadratic polynomial fitting is used, and error bars are smaller
than symbols.
}
\label{fig:su6_dimer}
\end{figure}

In this part, we show that although the N\'{e}el order is absent in
the honeycomb lattice in both the SU(4) and SU(6) case,
the VBS order appears after the single-particle gap opens.

The microscopic mechanisms of the VBS and N\'{e}el orders in the
honeycomb lattice are different from those in the square lattice.
In the square lattice, the N\'{e}el wave vector $(\pi,\pi)$ matches the
Fermi surface nesting condition, while the cVBS wave vectors at
$(\pi,0)$ or $(0,\pi)$ do not.
This explains why the N\'{e}el order wins over the VBS order in
the weak and intermediate coupling regime for the SU(4) and SU(6) cases.
The dimer ordering could win in the deep Mott insulating states
in which the local moments rather than Fermi surfaces play the
leading role.
In contrast, in the honeycomb lattice the pVBS and cVBS orders break
the translational symmetry.
Their wave vectors connect two different Dirac points $\vec K$ and
$\vec K^\prime$ and can be viewed as a nesting between Fermi points.
The N\'{e}el order does not break translational symmetry and generates
gap within each Dirac point.
Hence, in the honeycomb lattice, the N\'{e}el order has no particular
advantage compared to the VBS order from the perspective of Fermi
surface nesting.

The finite-size scalings of the VBS order parameter $\mbox{dim}_K(L)$
of the SU(4) case are presented in Fig. \ref{fig:su4_dimer}.
The long range VBS order starts to appear at $6<U<8$ as shown
in Fig. \ref{fig:su4_dimer} ($a$), which is consistent with
the single-particle gap opening at $U_c\approx 7$.
As further increasing $U$, the VBS order parameter in the
$L\to\infty$ limit becomes non-monotonic:
it keeps increasing until reaches the maximum around $U\approx 10$.
After that, it begins to decrease as further increasing $U$ as shown
in Fig. \ref{fig:su4_dimer} ($b$) for $U\ge 12$.
In this case, the extrapolated values of $\mbox{dim}_K(L)$ at
$1/L\to 0$ are very weak.
It is difficult to judge whether the VBS order vanishes.
Nevertheless, the spin gap data in Fig. \ref{fig:spingapL} ($a$)
still show noticeable values, and
thus are consistent with a weak VBS order.

The analysis for the SU(6) case can be carried out in parallel.
The finite-size scalings of $\mbox{dim}_K(L)$ are presented in
Fig. \ref{fig:su6_dimer} ($a$) and ($b$).
The long-range order starts to appear at $U=10\sim 11$, which is
consistent with the critical value of $U_c$ for opening the
single particle gap.
Similar to the SU(4) case, the VBS order parameter in the $L\to\infty$ limit
behaves non-monotonically as increasing $U$.
The overall scale of the VBS order parameter in the SU(6) case
is larger than that in the SU(4) case because the VBS order
is strengthened by increasing $2N$.

\begin{figure}[htb]
\includegraphics[width=\linewidth]{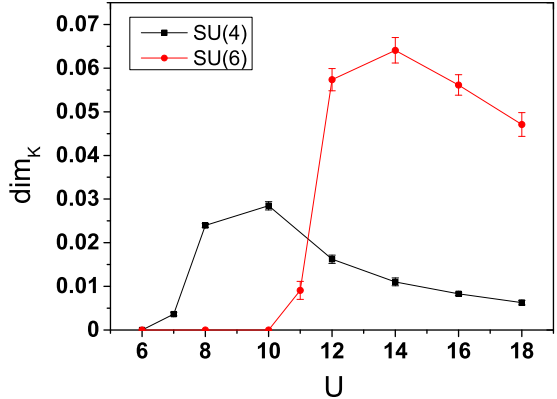}
\caption{The extrapolated VBS order parameter $\mbox{dim}_K$
as a function of $U$ for the SU(4) and SU(6) cases.
The critical values of $U_c$ for the appearances of the VBS order
are consistent with those of the single-particle gap $\Delta_{sp}$
opening shown in Fig.~\ref{fig:single_gap}.
}
\label{fig:GSdiag}
\end{figure}

Fig. \ref{fig:GSdiag} presents the extrapolated VBS order parameter as
a function of $U$.
With increasing $U$, the system undergoes a phase transition from the
Dirac semi-metal phase to the VBS phase.
The reason of the non-monotonic behavior of the VBS order parameter
at  $U>U_c$ is the following.
The VBS order is essentially the spatial variation of the kinetic
energy of each bond.
For $U\approx U_c$, the system remains in the intermediate coupling regime.
A dimerized bond in this regime is an SU($2N$) singlet state,
but it is not the singlet based on the Heisenberg model
Eq. \ref{eq:superexchange}: Each site has significant charge fluctuations,
and the kinetic energy scale on each bond is at the order of $t$.
In contrast, in the large-$U$ limit, charge fluctuations are suppressed.
The kinetic energy contributes through the 2nd order perturbation
process, i.e., the super-exchange effect described by
Eq. \ref{eq:superexchange}.
The overall energy scale of the bond kinetic energy is suppressed to
the order of $t^2/U$.
Thus after the initial increase of the VBS order just after
$U>U_c$, further increasing $U$ reduces the overall kinetic
energy scale, which suppresses the VBS order strength.

\subsection{Nature of the VBS ordering pattern}
\label{sect:p_or_c}

\begin{figure}[htb]
\includegraphics[width=\columnwidth]{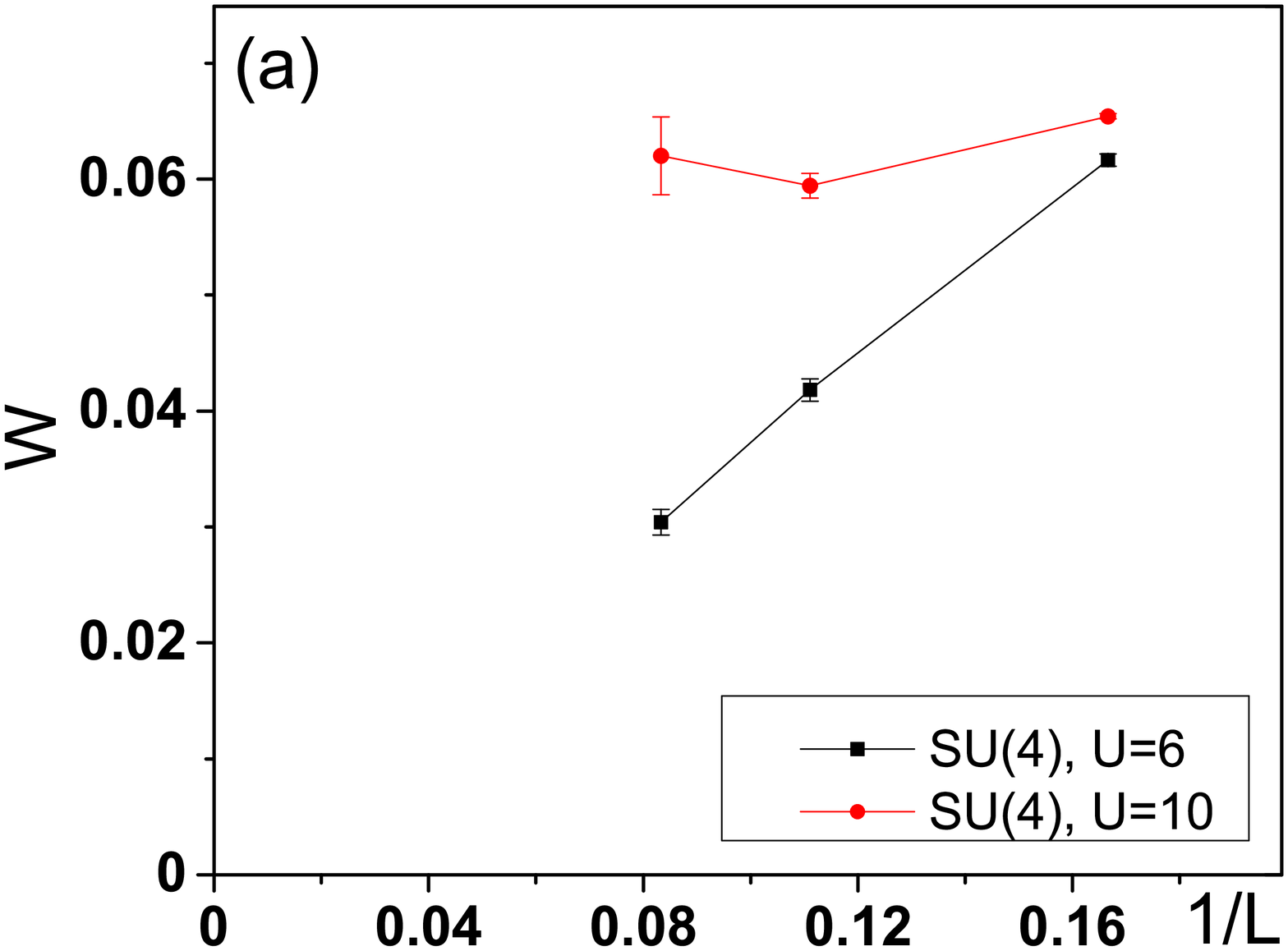}
\includegraphics[width=\columnwidth]{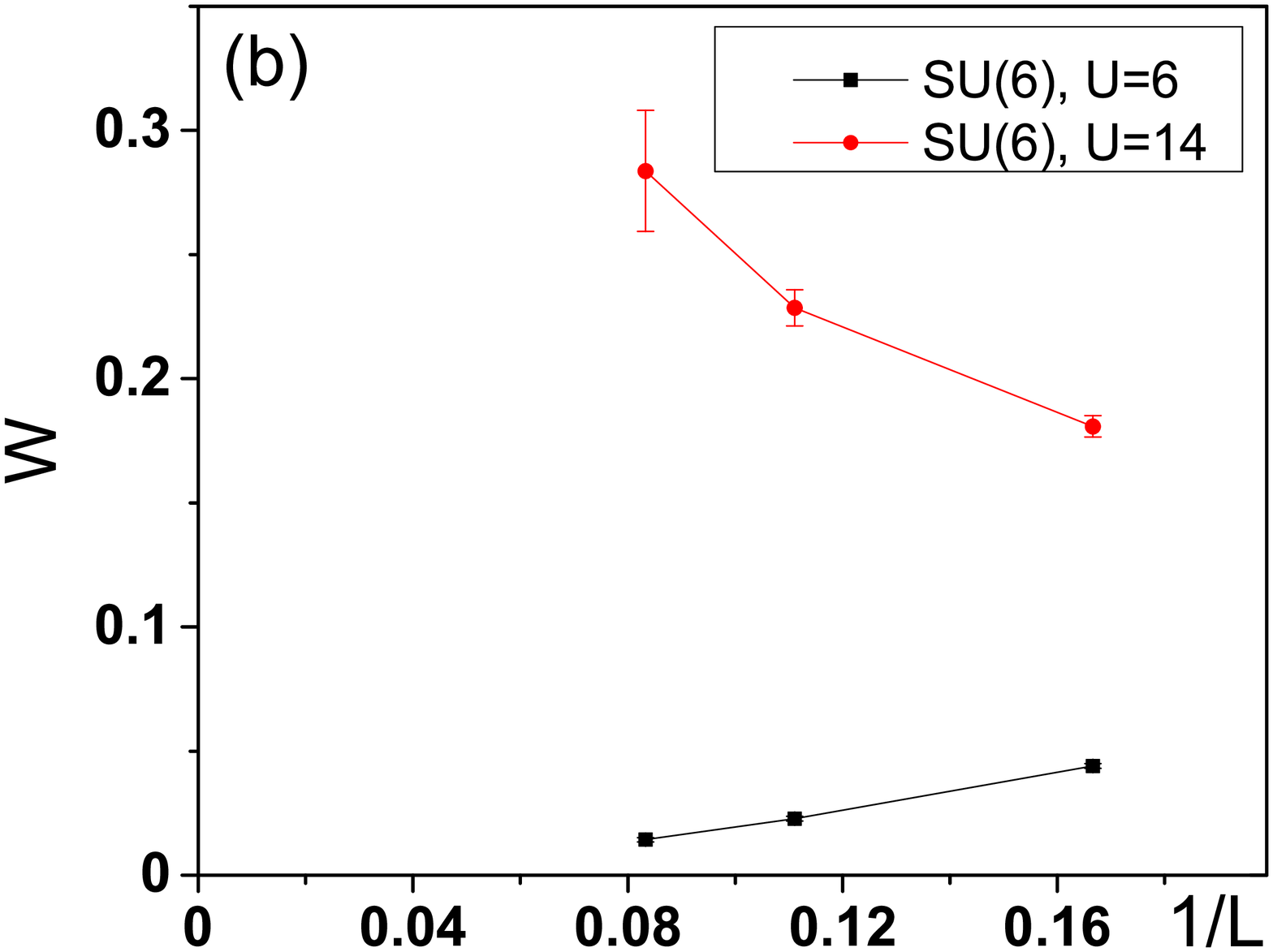}
\caption{ Finite size scalings of $W$ with different parameters $U$
and $2N$: ($a$) SU(4) case with $U=6,10$, and ($b$) SU(6) case
with $U=6,14$.
The results clearly point out the VBS order at $U>U_c$ is cVBS.
}
\label{fig:W_dimer}
\end{figure}

As we explained in Sect. \ref{sect:model_order}, both the
cVBS and pVBS configurations exhibit the same symmetry breaking pattern.
They are represented by the same complex order parameter $D_{K,1}$,
and cannot be distinguished from the structural factor scalings
which only yield magnitude of $D_{K,1}$.
Nevertheless, the distributions of the argument angles
are different between the cVBS and pVBS orders.
We investigate the nature of the VBS order by calculating $W$
following the definition in Eq. \ref{eq:histW}.

In Fig. \ref{fig:W_dimer} ($a$) and ($b$), we present the finite size
scalings of $W$ for the SU(4) and SU(6) cases, respectively.
As mentioned below Eq. \ref{eq:histW}, $W=1$ for a classic cVBS
configuration, $W=-1$ for a classic pVBS configuration, and
zero in the Dirac semi-metal phase.
The finite size scalings of $W$ indicate the most probable type of the
VBS configuration.
In both SU(4) and SU(6) cases, $U=6$ lies in semi-metal phase,
and thus $W$ drops as increasing the lattice size.

For both cases of SU(4) with $U=10$ and SU(6) with $U=14$,
the systems are in the VBS order phases.
The finite-size scalings of $W$ in Fig.~\ref{fig:W_dimer} ($a$) and
($b$) show that they saturate to positive values indicating the
cVBS instead of pVBS configuration.
The values of $W$ in the SU(6) are about one order higher
than those in the SU(4) cases, indicating much stronger VBS order.

\section{Absence of the current order}
\label{sect:current}

Besides the N\'{e}el and the VBS dimer orders discussed in the main text,
a spontaneous current order provides another possibility to open an
energy gap for the SU($2N$) Dirac fermions.
The effect of such current order (or loop currents) is similar to
those give rise to the topological band structure in the spinless
Haldane model \cite{Haldane1988}, or, spin-1/2 Kane-Mele
model \cite{Kane2005}. In this section, we present
the simulations on current orderings.
\begin{figure}[htb]
\includegraphics[width=0.5\columnwidth]{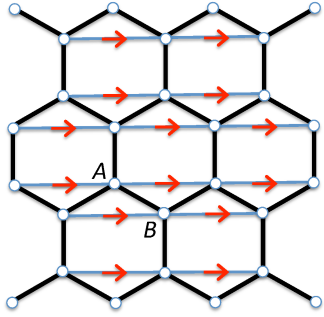}
\caption{The positive direction of current order on the honeycomb lattice.}
\label{fig:current_direction}
\end{figure}

\begin{figure}[htb]
\includegraphics[width=0.9\columnwidth]{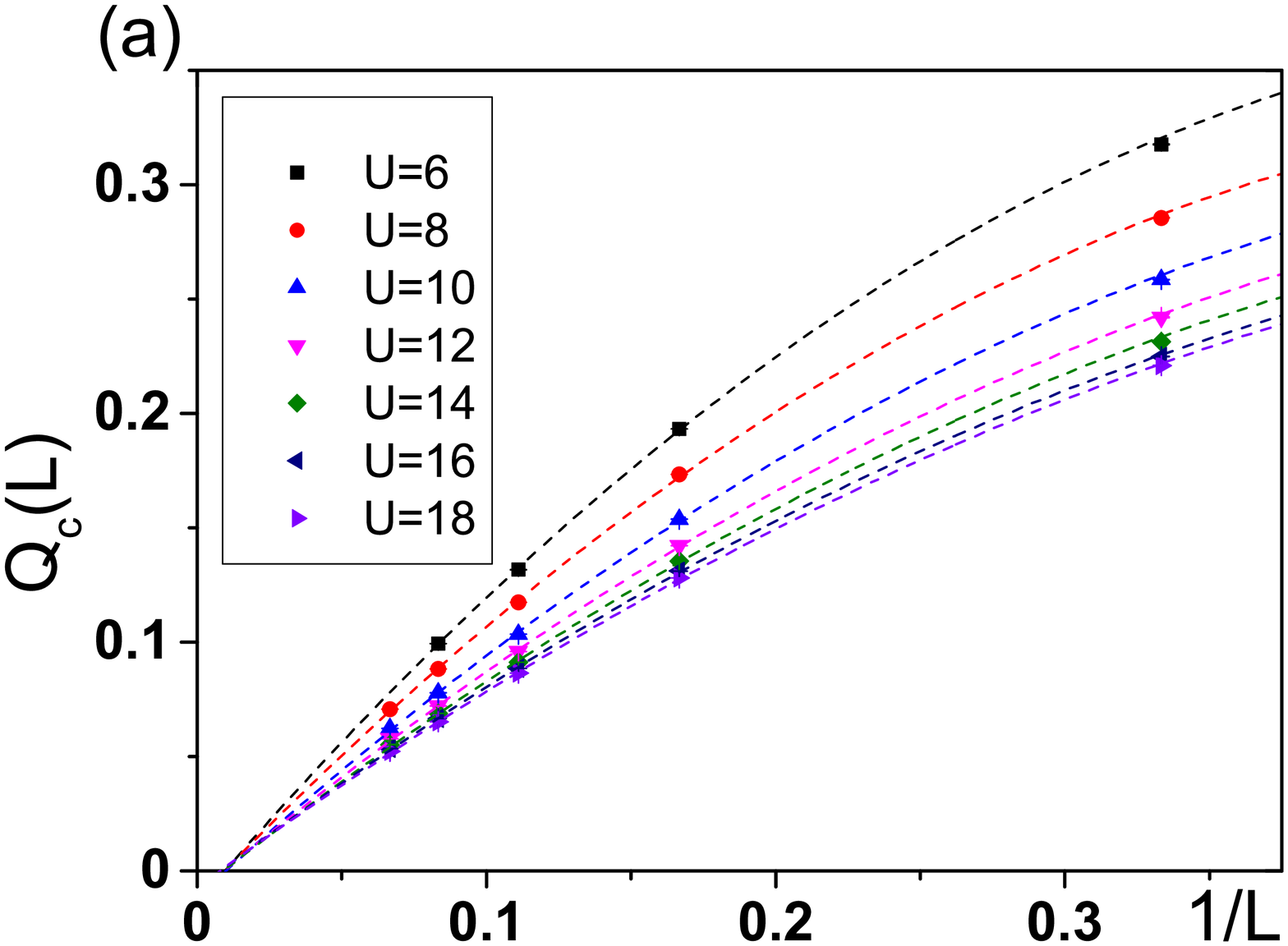}
\includegraphics[width=0.9\columnwidth]{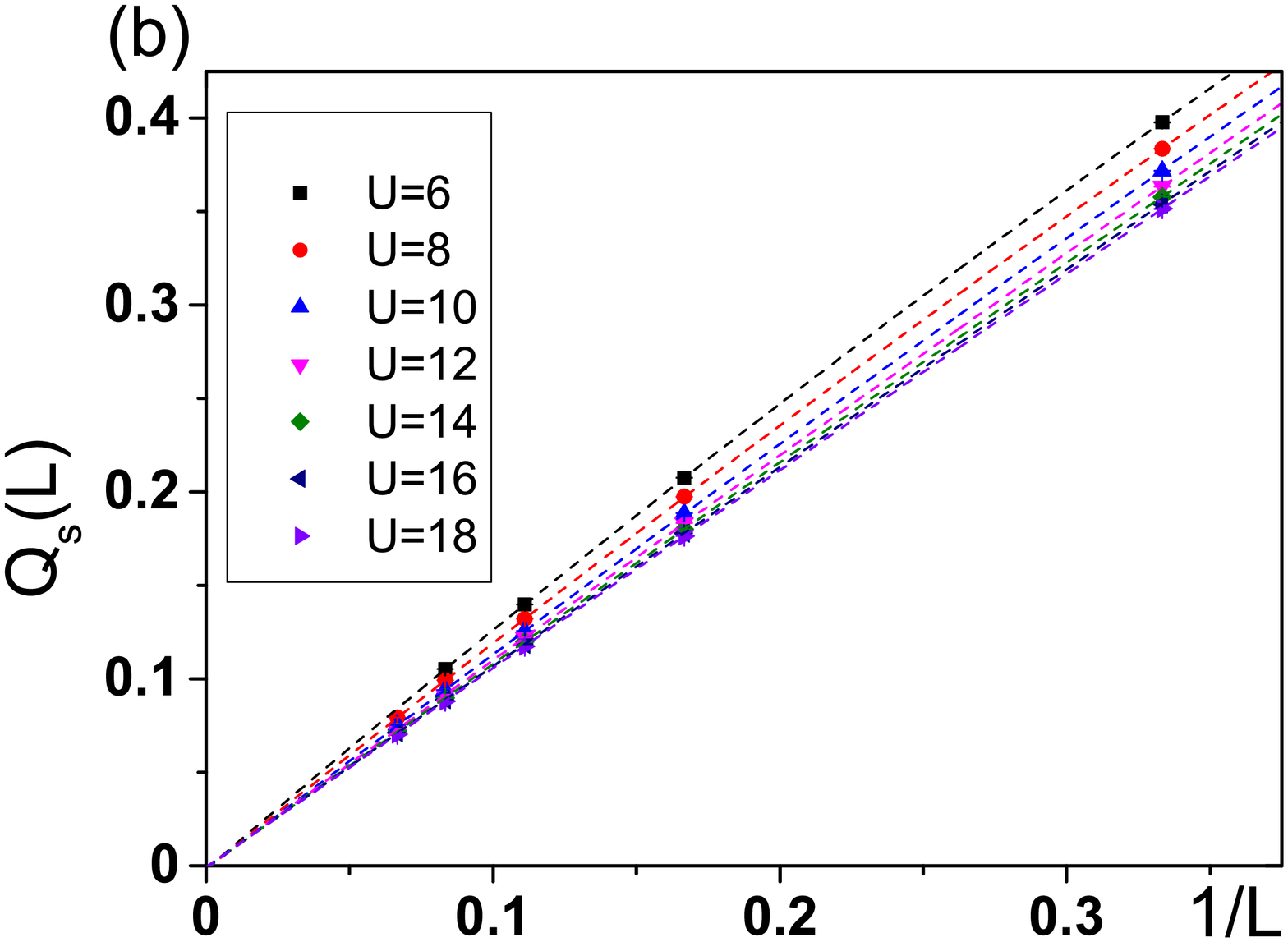}
\caption{The finite size scalings of the charge loop current
order $Q_{c}(L)$ and spin loop current order $Q_{s}(L)$
for the SU($4$) case.
The quadratic polynomial fitting is used.
Error bars of QMC data are smaller than symbols.}
\label{fig:curt_su4}
\end{figure}
\begin{figure}[htb]
\includegraphics[width=0.9\columnwidth]{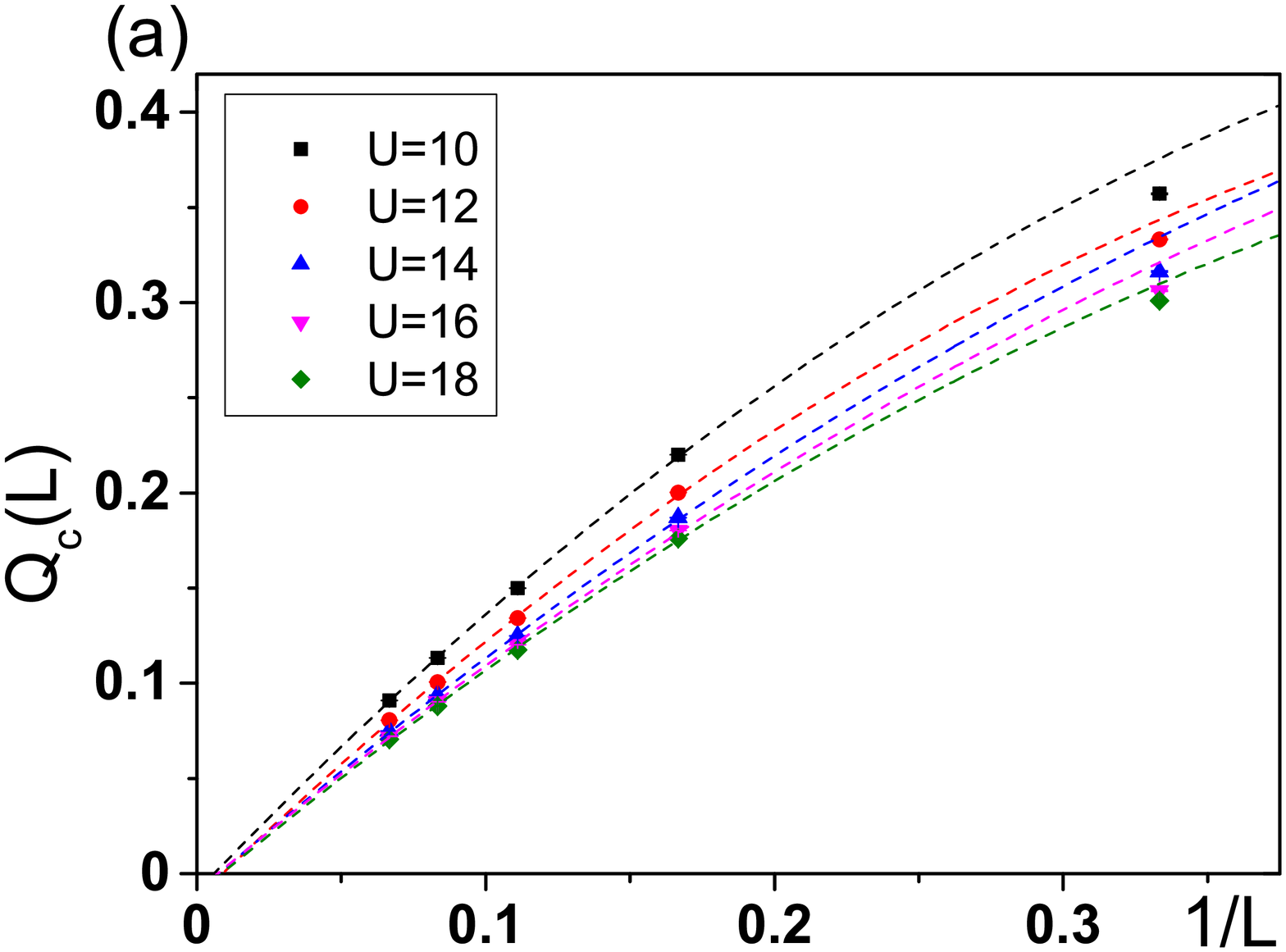}
\includegraphics[width=0.9\columnwidth]{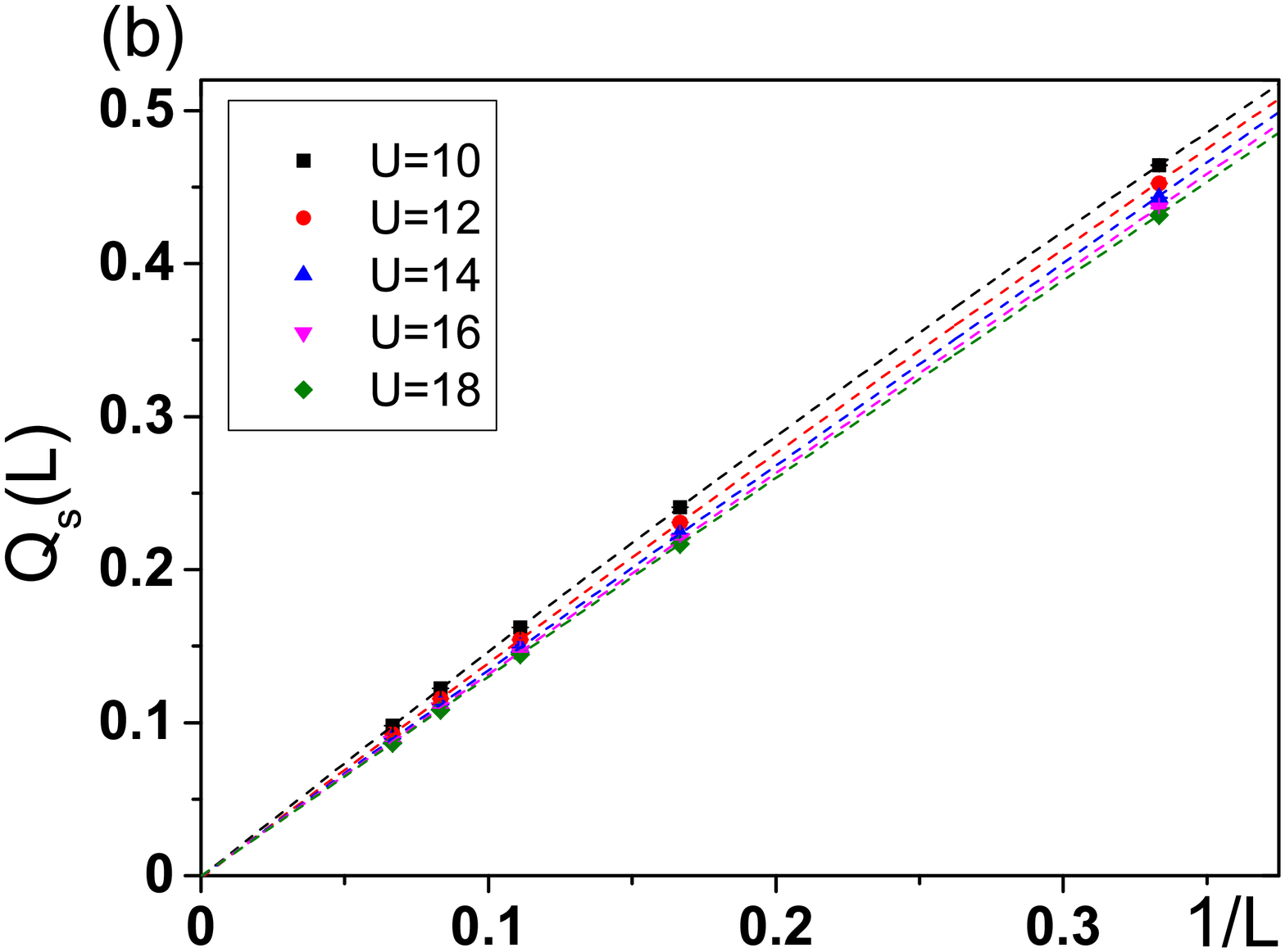}
\caption{The same as Fig.~\ref{fig:curt_su4} but for the case of
SU(6) and with different values of $U$.}
\label{fig:curt_su6}
\end{figure}

We define the following current operator for each fermion component
$\alpha$ as
\bea
J^\alpha_{jj'}=i\left[\left(c_{j,\alpha}^{\dagger}
c_{j',\alpha}-c_{j',\alpha}^{\dagger}c_{j,\alpha}\right)\right],
\eea
where $jj'$ represent the next-nearest-neighbors
and the positive direction of currents follows the arrows
indicated in Fig. \ref{fig:current_direction}.
For simplicity, we only consider the horizontal bonds.
Furthermore, following Ref. [\onlinecite{zheng2011}],
the charge and spin current order parameters are defined as
\bea
J^{c(s)}_{jj^\prime}=\sum_\alpha F(\alpha) J^a_{jj^\prime},
\eea
where $F(\alpha)=\pm 1$ decides the current direction of
the spin component $\alpha$.
We set $F(1,...,2N)=+1$ for $J^c$, while for spin current,
$F(1,...,N)=+1$ and $F(N+1,...,2N)=-1$ without loss of generality.
Then the loop current structure factors are defined as
\bea
Q_{c(s)}=\lim_{L\rightarrow \infty}\frac{1}{2L^{2}} \sqrt{\sum_{jk}(-1)^{j+k}J^{c(s)}_{jj'}
J^{c(s)}_{kk'}}.
\eea

The simulation results are shown in Fig.~\ref{fig:curt_su4}
and Fig.~\ref{fig:curt_su6} for the SU(4) and SU(6) cases,
respectively.
In both cases, both long-range charge and spin loop current orders
are absent in the thermodynamic limit, excluding the
possibility of spontaneous current orders as the cause of
the gap opening in both the SU(4) and SU(6) cases.

\section{Discussion on the nature of the Dirac-to-cVBS transition}
\label{sect:transition}

From the above QMC analysis, we have identified the quantum phase
transition between the Dirac semi-metal phase and the cVBS
phase as $U$ increases in both SU(4) and SU(6) cases.
As discussed previously, the order parameter $D_{\vec K,1}$ unifies
two different VBS patterns: the cVBS and pVBS, which
correspond to positive and negative values of $D_{\vec K,1}$, respectively.
This means that positive and negative values of $D_{\vec K,1}$ are
non-equivalent to each other.
As a result, in principle, the cubic order terms $D_{\vec K,1}^3$ and
$D_{\vec K,1}^{3,*}$ are allowed by symmetry in the effective action
of the VBS state.
Based on the Ginzburg-Landau (GL) theory, such odd order terms generally
lead to first order phase transitions.
However, we do not find strong evidence of first order phase transitions
in our QMC simulations.
This may be caused by a suppression of the third order terms by the coupling
to the critical fluctuations of Dirac fermions close to the
quantum phase transition point as analyzed below.

\subsection{The Ginzburg-Landau analysis}

\begin{figure}[h]
\includegraphics[width=0.4\textwidth]{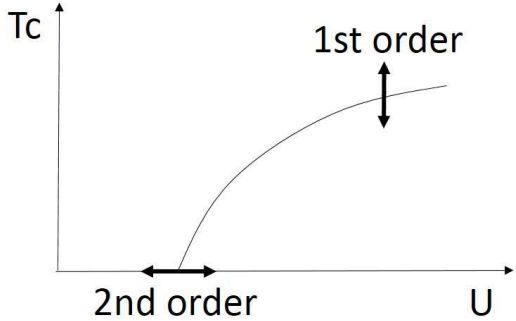}
\caption{Putative phase diagram of the Dirac semi-metal-to-
cVBS phase transition on the honeycomb lattice.
The zero temperature transition can be of the 2nd order
due to the coupling to the gapless Dirac fermions,
while the finite temperature transition should
generally be of the 1st order.
}
\label{fig:GLphase}
\end{figure}

In order to describe the Dirac semi-metal-to-VBS transition, we
construct the following GL free energy density $f(\psi,\psi^*)$ where
$\psi$ is the gap function associated with the cVBS order
$D_{\vec K, 1}$ which carries the energy unit.
$f$ does not possess the symmetry $\psi\to -\psi$, but needs to
maintain the lattice translation symmetry.
Based on the fact that $\psi$ and $\psi^*$ carry momenta $\pm \vec K$,
respectively,  the analytic part of the GL free energy is constructed
as
\begin{eqnarray}
f_A(\psi, \psi^*)=r_2|\psi|^2 - r_3(\psi^{*3}+\psi^3)+r_4|\psi|^4.
\label{eq:free_analytic}
\end{eqnarray}
The appearance of the $r_3$-terms is due to the lattice momentum conservation
$3 \vec K\equiv 0$ in the honeycomb lattice.
The $r_3$-terms ensure the non-equivalence between cVBS ($\psi^3>0$)
and pVBS ($\psi^3<0$).
Based on the QMC results, the cVBS state wins over the pVBS state,
thus $r_3>0$.
According to the GL theory, the $r_3$-terms lead to the 1st order
phase transition.
At $r_{2t}=r_3^2/r_4$, there exist 4-degenerate energy minima located
at $\psi=0$ and $\psi=\frac{r_3}{r_4} e^{i\theta}$
with $\theta=0,\pm \frac{2}{3}\pi$, respectively.
The former minimum corresponds to disordered state, while
the latter three minima correspond to the cVBS state.
As $r_2$ is lowered below $r_{2t}$, the ground state configuration
changes from the disordered semi-metal phase to the cVBS state
discontinuously.

The above 1st order phase transition could be weak if $r_3/r_4\ll 1$.
A continuous phase transition is recovered in the limit of $r_3/r_4\to 0$.
The value of $r_3$ and $r_4$ can be estimated by considering their
physical processes as follows.
Since $\psi$ carries momentum $K$, the dominant
contribution to $r_3$ comes from the scattering processes of
$K \rightarrow K' \rightarrow \Gamma \rightarrow K$
and $K \rightarrow \Gamma \rightarrow K' \rightarrow K$
where $K, K^\prime$ and $\Gamma$ represent small
regions centered around these momenta, respectively.
The involvement of the high energy point $\Gamma$ strongly reduces
the value of $r_3$, which should be proportion to the inverse
of the band width at the scale of $1/t$.
On the other hand, with similar analysis to the $r_4$ term,
the dominant scattering processes correspond to $K\rightarrow
K'\rightarrow K\rightarrow K'\rightarrow K$, all of which
are in the low energy region.
Therefore, $r_3/r_4$ is expected to be small.

So far, we only consider the analytic part of the GL free energy in
Eq. \ref{eq:free_analytic}.
Due to the coupling between $\psi$ with the gapless fermions, the free
energy potentially contains a non-analytic part even at the
mean-field level as analyzed below.
In our system, there exist $4N$ low energy Dirac cones.
After the developing of the cVBS order, the single particle
spectrum around each Dirac point at the mean-field level
becomes $E_k=\sqrt {v^2 k^2 +|\psi|^2}$,
where $\vec k$ is the deviation from the location of the
Dirac point.
We can estimate the free energy density at the mean-field
level arising from the low energy spectra around the Dirac points as
\bea
f_L&\approx& -\frac{4N}{\beta} \int_0^\Lambda
\frac{d^2\vec k }{(2\pi)^2}
\left(\ln (1+e^{\beta E_k})+\ln (1+e^{-\beta E_k}) \right) \nn \\
&+&\frac{1}{ 2N g} |\psi|^2,
\label{eq:free_L}
\eea
where $\Lambda$ is the momentum cut-off, and $\beta$ is
the inverse of temperature.
If taking the zero temperature limit first, i.e., $\beta\to \infty$,
and then taking the limit of $|\psi|\to 0$,
after performing the integral, we arrive at a non-analytic
part not included in Eq. \ref{eq:free_analytic},
\bea
f_{n}=r_{3,n} |\psi|^3,
\eea
where $r_{3,n}=\frac{2N}{3\pi v^2}>0$.
If $r_{3,n}>r_3$, then $r_{3,n}$ and $r_3$-terms combine together
are positive definite, which cannot induce 1st order
transition.
If $r_{3,n}<r_3$, the first order phase transition remains
but is weakened.

Since the semi-metal-to-cVBS transition only breaks discrete symmetry,
it is expected to survive at finite temperatures.
There exists an additional contribution at a finite temperature
from Eq. \ref{eq:free_L}, which can be organized as
\bea
f_L (T) -f_L(T=0) =-\frac{4N}{\pi \beta^3 }
\int^{+\infty}_{\beta |\psi|}  y \ln (1+e^{-y}) dy,
\nn \\
\eea
where the upper limit of the integrand is set to $+\infty$.
If we take the limit of $|\psi|\to 0$ first and then set
$\beta$ at an arbitrarily low but still finite temperature,
the above expression contributes an extra non-analytic term
\bea
\Delta f_n(T)= -r_{3,n} |\psi|^3,
\eea
which precisely cancels $f_n$ at zero temperature.
As a result, the non-analytic part of the free-energy disappears at
finite temperature.
Thus finite temperature transition from the Dirac semi-metal-to-cVBS
still remains the 1st order.

The above analysis shows that the low energy
Dirac fermions can significantly changes the nature of the quantum
phase transition through non-analytic contributions to
the G-L free energy.
Nevertheless, this effect can only exist at zero temperature quantum
phase transition, not at finite temperature phase transition,
as summarized in Fig. \ref{fig:GLphase}.
In Sect. \ref{subsect:mean-field}, we perform a numeric study
based on the mean-field theory, which agrees with the
above analytic results.

Nevertheless, the above analysis is only at mean-field level.
The strong quantum fluctuations due to the coupling between
the VBS order and the gapless Dirac fermions may
further soften the 1st order transition and drive the
transition to be continuous.
If this is true, it means that in the correlated SU($2N$) Dirac fermion
systems, an exotic continuous quantum phase transition beyond typical
GL paradigm is realized.
Within the accuracy of the current QMC simulations, we cannot judge the
nature of the  Dirac semi-metal-to-cVBS transitions at the SU(4)
and SU(6) cases are of weak 1st order or continuous 2nd order.
We leave the further theoretical and numerical analyses to future works,
but hope the conclusion by now is strong enough to motivate ultra-cold
atom experiments to realize such exotic quantum phase transition.

\subsection{A Mean-field theory calculation}
\label{subsect:mean-field}
\label{subsect:mean-field}
\begin{figure}[h] \includegraphics[width=0.5\textwidth]{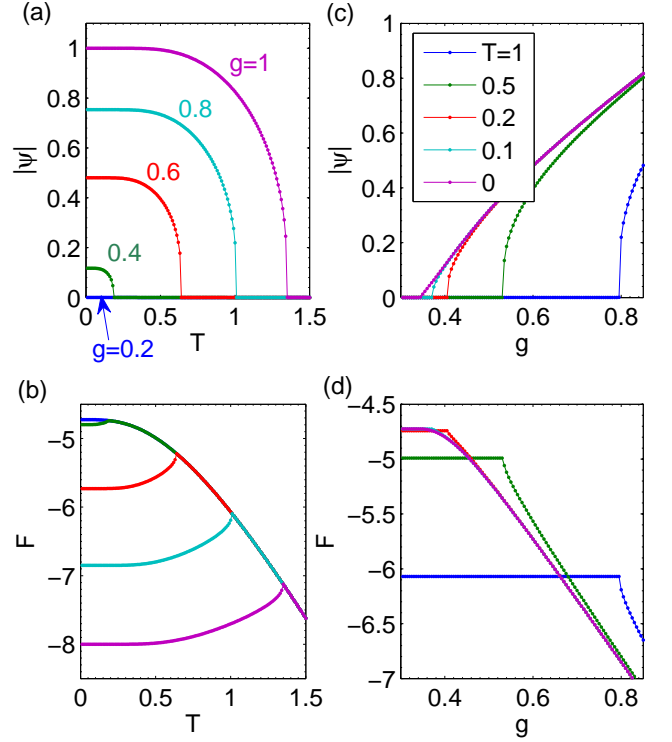}
\caption{Self-consistent mean field calculation results.
In (a) and (b), the VBS order parameter $|\psi|$ and free energy $F$
are plotted as a function of the temperature $T$. In (c) and (d),
$|\psi|$ and $F$ are plotted vs the effective interaction $g$. Linear
lattice size $L=99$ is used for finite $T\geq 0.01$, while
larger $L=300$ is used in the $T=0$ case in order to eliminate the finite size
effect.}
\label{fig:meanfield} \end{figure}

In this part, we present a numeric solution to the mean-field theory
at both finite and zero temperatures to illustrate the nature
of the above transitions.
Instead of directly using the Hubbard model, we employ a
phenomenological model exhibiting the order parameter
$D_{\vec K,1}$ in the cVBS channel with the following interaction
term
\begin{eqnarray}
H_I=-g\sum_{i}D_{i,1}^\dag D_{i,1},
\end{eqnarray}
where $g>0$ is the effective interaction and
$D_{i,1}=D_{\vec K,1}\exp(i\vec K \cdot r_i)$.
In the mean-field approximation,
\begin{eqnarray}
H_I\approx -\sum_{i}\left[\left(D_{i,1}^\dag
\psi \mathrm{e}^{i\vec K\cdot r_i} + h.c.\right) - \frac{1}{g}\psi^*\psi\right] ,
\end{eqnarray}
where $\psi=g\langle D_{\vec K,1} \rangle$.

After Fourier transformation, the mean field Hamiltonian is
\begin{eqnarray}
H_{MF}=\sum_{\vec k} \left[\left(\Psi_{\vec k,a}^\dag h_{\vec k}
\Psi_{\vec k,b} + h.c.\right) + \frac{1}{g}\psi^*\psi\right] ,
\end{eqnarray}
where $\Psi_{\vec k,a|b}=[c_{\vec k,a|b},c_{\vec k+\vec K,a|b},
c_{\vec k-\vec K,a|b}]^t$ and the matrix kernel $h_{\vec k}$ is
\begin{eqnarray}
h_{\vec k}=\left[
\begin{array}{ccc}
f_{\vec k}&-\psi f_{\vec k- \vec K}&-\psi^* f_{\vec k+\vec K}\\
-\psi^* f_{\vec k-\vec K}&f_{\vec k+\vec K}&-\psi f_{\vec k}\\
-\psi f_{\vec k+\vec K}&-\psi^* f_{\vec k}&f_{\vec k-\vec K}
\end{array}
\right] ,
\end{eqnarray}
in which $f_{\vec k}=\sum_{j} \exp(i\vec k\cdot \hat{e}_j)$.
The value of $\psi$ can be solve self-consistently through definition
\begin{eqnarray}
\psi=\frac{g}{L^2}\sum_{\vec k} \left[ f_{\vec k-\vec K} \langle c_{\vec k+\vec K,a}^\dag c_{\vec k,b} \rangle + f^*_{\vec k-\vec K} \langle c_{\vec k+\vec K,b}^\dag c_{\vec k,a} \rangle \right] , \nonumber \\
\end{eqnarray}
where $\exp(i\vec K\cdot\hat{e}_j)=\{1,\omega,\omega^*\}$ has been used to arrive at the above self-consistent relation.

The self-consistent results are shown in Fig.~\ref{fig:meanfield}.
At finite temperature $T>0$, the jumps of $|\psi|$ at critical points
$T_c$ and $V_c$ indicate a first order phase transition, which
is more obvious as shown by the cusps in the free energy data
at transitions.
However, as temperature decreases, the free energy as a function
of $V$ exhibits more smooth behavior, as shown in Fig.~\ref{fig:meanfield}(d).
Meanwhile, the discontinuous behavior $|\psi|$ at $V_c$ becomes weaker
and weaker and finally vanishes, resulting in a continuous quantum
phase transition at zero temperature.

\section{Conclusions}
\label{sect:concl}
Based on all the simulations and analysis above, we present a ground-state
phase diagram of the SU($2N$) Hubbard model on the honeycomb
lattice with $2N=4,6$.
Quantum phase transitions from the Dirac semi-metal phases to the Mott
insulating phases with increasing $U$ are realized.
The transition values of $U_c$ increase with the number of
fermion components $2N$.
The Mott-insulating phases exhibit the cVBS order in the SU(4) and SU(6)
cases in contrast to the antiferromagnetic N\'{e}el order
previously studied in the SU(2) case \cite{Meng2010,Sorella2012,Assaad2013}.
The VBS signal is weak for SU($4$) comparing with SU($6$).
Close to $U_c$, the cVBS order increases sharply, and it is
suppressed as $U$ further increases, due to the suppression
of the overall kinetic energy scale in the strong
Mott-insulating region.
The nature of the semi-metal-to-cVBS transition is analyzed
at the mean-field level, which can be of the 2nd order
at zero temperature due to coupling to gapless Dirac
fermions but still remains at the 1st order at finite
temperatures.

{\it Note added-}
Upon finishing the revised (2nd) version, we became aware of the
interesting work by Li {\it et. al} \cite{Li2015} in which
the continuous quantum phase transitions from the semi-metal phase
to the cVBS phase are proposed and analyzed.

\acknowledgments
Z. Z., Y. W., and C. W. gratefully acknowledge financial support from the National Natural
Science Foundation of China under Grant Nos. 11574238 and 11328403, and the Fundamental Research Funds for the
Central Universities. D. W. acknowledges the support from National Natural Science
Foundation of China (11504164).
C. W. is supported by the NSF DMR-1410375 and AFOSR FA9550-14-1-0168.
C. W. acknowledges the support from the Presidents
Research Catalyst Awards of University of California.
Z. Y. M. is supported by the National Natural Science Foundation
of China (Grant Nos. 11421092 and 11574359) and the National
Thousand-Young-Talents Program of China.

\appendix

\section{Exact discrete Hubbard-Stratonovich transformation for the SU(4)
and SU(6) Hubbard interactions}
\label{append:HS}

For the SU(2) Hubbard model, the discrete HS decomposition can be performed
exactly\cite{hirsch1983}.
For more complicated interactions and more flavors of spins
such as in the SU($2N$) $t$-$J$ or Hubbard models, the interaction terms are
often decomposed into a set of discrete Ising fields with an systematic
error at the order of $(U\Delta \tau)^4$ \cite{Assaad2005,Assaad1998}.
This can be improved by an exact HS transformation for the SU($2N$)
Hubbard model studied here, which is particularly useful
at large values of $U$.

According to Ref.~\onlinecite{Wang2014}, the exact HS transformation
for SU(4) and SU(6) Hubbard interaction is expressed as
\begin{equation}
e^{-\frac{\Delta\tau U}{2}(n_{j}-N)^{2}}=\frac{1}{4}\sum_{l=\pm1,\pm2}
\gamma_{j}(l)e^{i\eta_{j}(l)(n_{j}-N)},
\label{eq:hs}
\end{equation}
which employs two sets of discrete HS fields $\gamma$ and $\eta$.
For the cases of $2N=4$ and $6$,  the Ising fields take values of
\begin{eqnarray}
\nonumber \gamma(\pm1)&=&\frac{-a(3+a^{2})+d}{d},\\
\nonumber \gamma(\pm2)&=&\frac{a(3+a^{2})+d}{d},\\
\nonumber \eta(\pm1)&=&\pm\cos^{-1}
\left\{ \frac{a+2a^{3}+a^{5}+(a^{2}-1)d}{4}\right\},\\
\nonumber \eta(\pm2)&=&\pm\cos^{-1}
\left\{ \frac{a+2a^{3}+a^{5}-(a^{2}-1)d}{4}\right\},
\end{eqnarray}
where $a=e^{-\Delta\tau U/2}$, $d=\sqrt{8+a^{2}(3+a^{2})^{2}}$.

\section{Zero temperature QMC and absence of the sign
problem at half-filling}
\label{append:sign}

In this appendix, we briefly introduce the zero temperature determinant
QMC algorithm~\cite{Assaad2008} and prove the absence of the sign
problem for the half-filled SU($2N$) Hubbard model
in bipartite lattices.

The Hamiltonian is composed of kinetic and interaction parts
\bea
H=H_{K}+H_{I}.
\eea
The free part reads
\bea
H_{K}=\sum_{i,j,\alpha}c_{i,\alpha }^{\dag }K_{i,j}c_{j,\alpha },
\label{eq:HK}
\eea
where $K$ is the kinetic energy matrix and $\alpha=1,\cdots,2N$.
The interaction part is
\bea
H_{I}=\frac{U}{2}\sum_{i}(n_i-N)^{2},
\label{eq:hhint}
\eea
where $n_{i}=\sum_{\alpha}c_{i\alpha}^\dag c_{i\alpha}$.

The expectation value of a physical observable operator $\hat{O}$ at
zero temperature is defined as
\bea
\langle \hat{O}\rangle =\frac{\langle \psi _{0}|\hat{O}|\psi _{0}\rangle }{%
\langle \psi _{0}|\psi _{0}\rangle }=\frac{\langle \psi _{T}|e^{-\Theta H}%
\hat{O}e^{-\Theta H}|\psi _{T}\rangle }{\langle \psi _{T}|e^{-2\Theta
H}|\psi _{T}\rangle },
\label{eq:expec}
\eea
where $|\psi _{0}\rangle $ is the ground state; $\Theta$ is a projection
parameter large enough to ensure the trial wave function $|\psi
_{T}\rangle $ is projected to the ground state $|\psi _{0}\rangle $.
In our QMC simulations, we rename $2\Theta=\beta$ due to the similarity
between projector QMC and finite temperature QMC algorithms.

Since $H_K$ and $H_I$ are non-commutative, we perform the second order
Suzuki-Trotter decomposition
\bea
\mathrm{e}^{-\Delta\tau(H_K+H_I)}=\mathrm{e}^{-\Delta\tau H_K/2}
\mathrm{e}^{-\Delta\tau H_I}\mathrm{e}^{-\Delta\tau H_K/2}+o[(\Delta\tau)^3], \nn\\
\label{eq:trotter}
\eea
to divide $\Theta$ into M slices with discrete time interval
$\Delta\tau=\Theta/M$.
Then for each time slice, the discretized HS transformation of
the interaction term Eq. \ref{eq:hhint} is performed in the density channel
the same as that in Eq. \ref{eq:hs}.
The imaginary time propagator, i.e., the projection operator, is
represented as
\bea
e^{-\Theta H} &=&\sum_{\{l\}}
\Big\{ U_{\{l\}}(\Theta,0)
\prod_{i,p}\frac{\gamma _{i,p}(l)}{4}e^{-i\eta _{i,p}(l)N}
\Big\},
\label{eq:imgprop}
\eea
where
\bea
U_{\{l\}}(\Theta,0)&=&\prod_{\alpha=1}^{2N}
\prod_{p=M}^{1}e^{-\Delta \tau
\sum_{i,j}c_{i\alpha }^{\dag }K_{ij}c_{j\alpha }}e^{i\sum_{i}c_{i\alpha }^{\dag }
\eta _{i,p}(l)c_{i\alpha}} .\nn
\eea
Here,
$\gamma_{i,p}(l)$ and $\eta_{i,p}(l)$ are the space-time discretized
HS fields defined in Eq. \ref{eq:hs} with $l$ taking values of $\pm 1,\pm 2$;
$\sum_{\{l\}}$ represents the summation over the spatial and temporal
configurations of the HS field;
$U_{\{l\}}(\Theta,0)$ is the propagation operator for the HS configuration
$\{ l \}$.

The trial wave function $|\psi _{T}\rangle $ is required to be a Slater
determinant, which we will specify later.
Substituting Eq.~\ref{eq:imgprop} into Eq.~\ref{eq:expec}, we obtain
\begin{widetext}
\bea
\langle \hat{O} \rangle &=&
\frac{\sum_{\{l\}} \Big\{ \langle \psi _{T} ~|U_{l}
(2\Theta ,\Theta ) ~ \hat{O} ~U_{l}(\Theta, 0)
|\psi _{T}\rangle \prod_{i,p}\gamma
_{i,p}(l)e^{-i\eta _{i,p}(l)N}\Big\} }{
\sum_{\{l\}}\Big\{ \langle \psi _{T}|U_{l}(2\Theta ,0)|\psi _{T}\rangle
\prod_{i,p}\gamma _{i,p}(l)e^{-i\eta _{i,p}(l)N}\Big\}
}
=\sum_{\{l\}}P_{\{l\}} ~ \langle\hat{O}\rangle _{\{l\}},
\eea
\end{widetext}
where $\avg{O}_{\{l\}}$ is the average value of $\hat O$ for the space-time
HS configuration $\{ l \}$ defined as
\bea
\avg{O}_{\{l\}}=\frac{\langle \psi
_{T}|U_{\{l\}}(2\Theta ,\Theta ) ~\hat{O}~ U_{\{l\}}(\Theta ,0)|\psi _{T}\rangle }
{\langle \psi _{T}|U_{\{l\}}(2\Theta ,0)|\psi _{T}\rangle },
\eea
and $P_{\{l\} }$ is the corresponding probability of
the HS field configuration $\{ l \}$ as
\bea
P_{\{l\} }&=&\frac{1}{Z}
 ~\langle \psi _{T}|U_{l}(2\Theta ,0)|\psi _{T}\rangle
\prod_{i,p}\gamma _{i,p}(l)e^{-i\eta _{i,p}(l)N
}. \nn \\
\eea
$Z$ is defined as
\bea
Z=\sum_{\{l\}} \langle \psi _{T}|U_{l}(2\Theta ,0)|\psi _{T}\rangle
\prod_{i,p}\gamma _{i,p}(l)e^{-i\eta _{i,p}(l)N} .
\eea
The summation over the HS configurations $\{l\}$ can be done by the Monte
Carlo method.

Next we prove the absence of the sign problem for the SU($2N$) Hubbard model
at half-filling in the zero temperature QMC method, {\it i.e.}, the
probability $P_{\{l\}}$ is positive-definite.
We factorize the trial wave function as
$|\Psi_T\rangle=\otimes_{\alpha=1}^{2N} |\psi_T^{N_\alpha}\rangle$,
where $|\psi_T^{N_\alpha}\rangle$ is a Slater-determinant state
for spin-$\alpha$ electrons with the particle number $N_\alpha$.
Then  $P_{\{l\}}$ reads as
\begin{widetext}
\bea
P_{\{l\}}
&=&\frac{1}{Z}
\left[\prod_{\alpha=1}^{2N} \langle \psi^{N_\alpha}_T|
\prod_{p=2M}^{1}e^{-\Delta \tau
\sum_{i,j}c_{i\alpha }^{\dag }K_{ij}
c_{j\alpha }}e^{i\sum_{i}\eta _{i,p}(l)( c_{i\alpha }^{\dag
}c_{i\alpha }-\frac{1}{2}) } ~|\psi^{N_\alpha}_T\rangle\right]
\left[\prod_{i,p}\gamma _{i,p}(l)\right] ,
\eea
\end{widetext}
where the HS fields $\gamma _{i,p}(l)$ given by Eq.~\ref{eq:hs} are
positive-definite.

Let us perform a particle-hole transformation only to
the spin-$\alpha=N+1,\cdots,2N$ component
\bea
c_{i\alpha }^{\dag }\rightarrow d_{i\alpha }=(-1)^{i}c_{i\alpha
}^{\dag },\, c_{i\alpha }\rightarrow d_{i\alpha }^{\dag
}=(-1)^{i}c_{i\alpha },
\eea
then the Slater-determinant state  $|\psi_T^{N_\alpha}\rangle$
changes to another Slater-determinant state of holes
with the hole number $N_{L}-N_\alpha$
denoted as $|\psi_T^{h,N_{L}-N_\alpha}\rangle$.
We arrive at
\begin{widetext}
\bea
P_{\{l\}}
&=&\frac{1}{Z}
\prod_{\alpha=1}^{N}
\langle \psi_T^{N_\alpha}|~
\prod_{p=2M}^{1}e^{-\Delta \tau
\sum_{i,j}c_{i\alpha }^{\dag }K_{ij}c_{j\alpha }}
e^{i\sum_{i}\eta _{i,p}(l)( c_{i\alpha }^{\dag
}c_{i\alpha }-\frac{1}{2}) }~ |\psi_T^{N_\alpha}\rangle
\nn \\
&\times&
\prod_{\alpha=N+1}^{2N}\langle \psi_T^{h,N_{L}-N_\alpha}|~
\prod_{p=2M}^{1}e^{-\Delta \tau
\sum_{i,j}d_{i\alpha }^{\dag }K_{ij}d_{j\alpha }}
e^{-i\sum_{i}\eta _{i,p}(l)( d_{i\alpha }^{\dag
}d_{i\alpha }-\frac{1}{2})}
~|\psi_T^{h,N_{L}-N_\alpha}\rangle \nn \\
&\times&\left[\prod_{i,p}\gamma _{i,p}(l)\right].
\eea
\end{widetext}

Now we add back the explicit form of the Slater-determinant states
$|\psi_T^{N_\alpha}\rangle$ and $|\psi_T^{h,N_{L}-N_\alpha}\rangle$ as
\bea
|\psi_T^{N_\alpha}\rangle&=&\prod_{j=1}^{N_{\alpha}}
\Big(\sum_{i=1}^{N_{L}}c_{i\alpha}^{\dag} Q^\alpha_{i,j} \Big) |0\rangle
=\prod_{j=1}^{N_\alpha}\Big(\vec{c}_\alpha^{\dag }Q^\alpha \Big)_{j}|0\rangle, \nn\\
|\psi_T^{h,N_{L}-N_\alpha}\rangle&=&
\prod_{j=1}^{N_{L}-N_\alpha}\Big(\sum_{i=1}^{N_{L}} d_{i\alpha}^{\dag} \tilde{Q}_{ij}^\alpha \Big)
|0\rangle_h \nn\\
{}&=&\prod_{j=1}^{N_{L}-N_\alpha}\Big(\vec{d}_\alpha^{\dag }\tilde{Q}^\alpha\Big)_{j}|0\rangle_h, \nn\\
\eea
where $|0\rangle$ and $|0\rangle_h$ are the particle vacuum and hole
vacuum states, respectively;
$N_{L}$ is the number of lattice sites;
$Q_\alpha$ is a $N_{L}\times N_\alpha$-dimensional rectangular matrix,
and $\tilde{Q}_\alpha$ is a $N_{L}\times (N_{L}-N_\alpha)$-dimensional matrix;
$\vec{c}_\alpha^\dagger$ and $\vec{d}_\alpha^\dagger$ are vector notations
for $c^\dagger_{i\alpha}$ and $d^\dagger_{i\alpha}$ with $i=1$ to $N_{L}$.

A Slater-determinant wave function has nice properties as
\bea
e^{\vec c^{\dag} M \vec c} \prod_{j=1}^{N_p}(
\vec c^{\dag }Q)_j|0\rangle
=\prod_{j=1}^{N_{p}} [\vec c^{\dag }e^{M} Q]_j|0 \rangle,
\eea
and
\bea
&&\langle 0|\prod_{j=1}^{N_p}(\vec c Q^\dag)_j
~ e^{\vec c^{\dag} M \vec c} ~
\prod_{j=1}^{N_p}(\vec c^{\dag }Q^\prime)_j|0\rangle\nn \\
&=&\det \left[ Q^\dagger e^{M} Q^\prime \right],
\eea
where $M$ is an $N_{L}\times N_{L}$ Hermitian matrix, or anti-Hermitian
matrix.
\begin{figure}[htb]
\includegraphics[width=0.9\columnwidth]{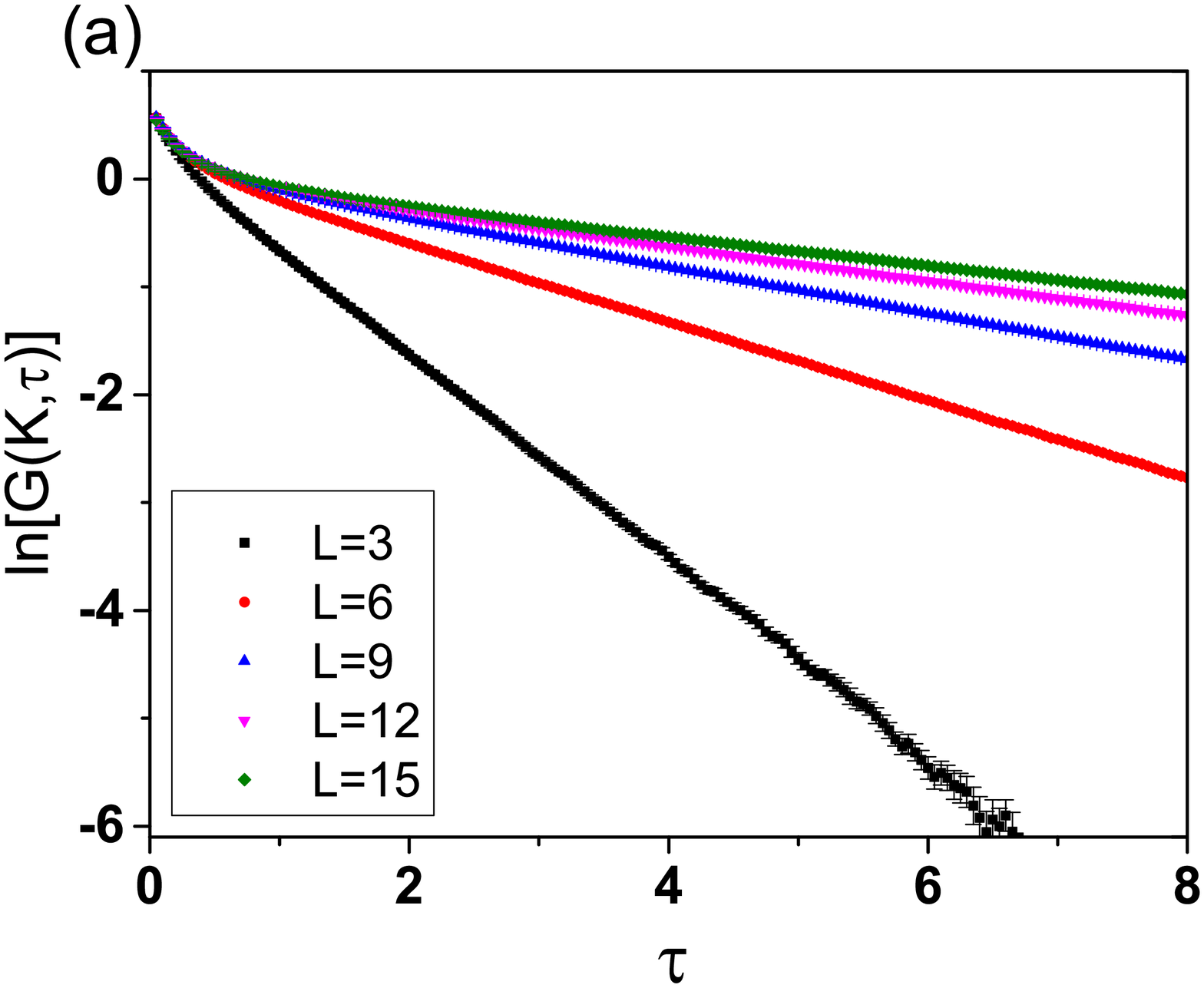}
\includegraphics[width=0.9\columnwidth]{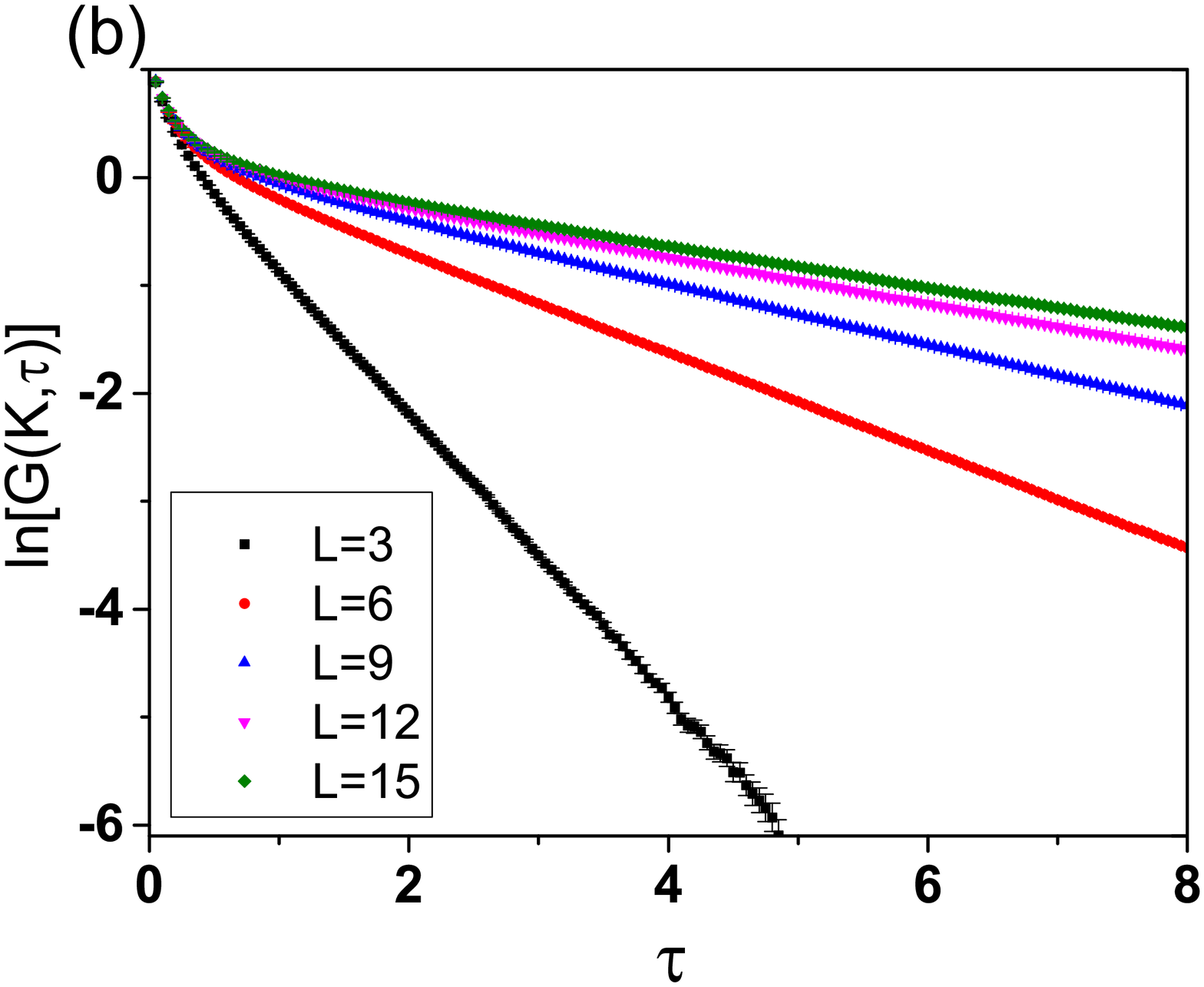}
\caption{$\ln G(K,\tau)$ as functions of $\tau$ at ($a$) $U=7$
for SU(4) and ($b$) $U=11$ for SU(6) at different values
of lattice size $L$.
The data are in good quality and the slopes (the $\Delta_{sp}$)
are reliably extracted.}
\label{fig:lnGK}
\end{figure}

Based on these properties, we have
\bea
P_{\{l \}}&=&\frac{1}{Z}\prod_{\alpha=1}^{N}
\det \left [ \left( Q_{\alpha }\right) ^{\dag }\left(
\prod_{p=2M}^{1}e^{-K}e^{iV_{p}(l)}\right) Q_{\alpha }\right]
\nn \\
&\times& \prod_{\alpha=N+1}^{2N}
\det\left[\left( \tilde{Q}_{\alpha }\right)^{\dag }\left(
\prod_{p=2M}^{1}e^{-K}e^{-iV_{p}(l)}\right) \tilde{Q}_{\alpha
}\right] \nn \\
&\times& \prod_{i,p}\gamma _{i,p}(l),
\eea
where
$V_{p}(l)$ is a purely real diagonal matrix whose $i$-th
diagonal element reads
\bea
[V_{p}(l)]_{ii}=\eta _{i,p}(l).
\eea

If we set the trial wave function to satisfy $N_{\alpha}=N_{L}/2$ and
thus $Q^{1,\cdots,N}=Q$, $\tilde{Q}^{N+1,\cdots,2N}=Q^*$, then we have
\bea
P_{\{l\}}&=&\frac{1}{Z} \left \vert \det \left[ Q
^{\dag }\left(
\prod_{p=2M}^{1}e^{-K}e^{V_{p}(l)}\right) Q\right ]  \right\vert ^{2N} \nn \\
&\times&
\prod_{i,p}\gamma _{i,p}(l),
\eea
thus the probability distribution $P_{\{l\} }$ is positive-definite
at half-filling.

\begin{figure}[htb]
\includegraphics[width=0.9\columnwidth]{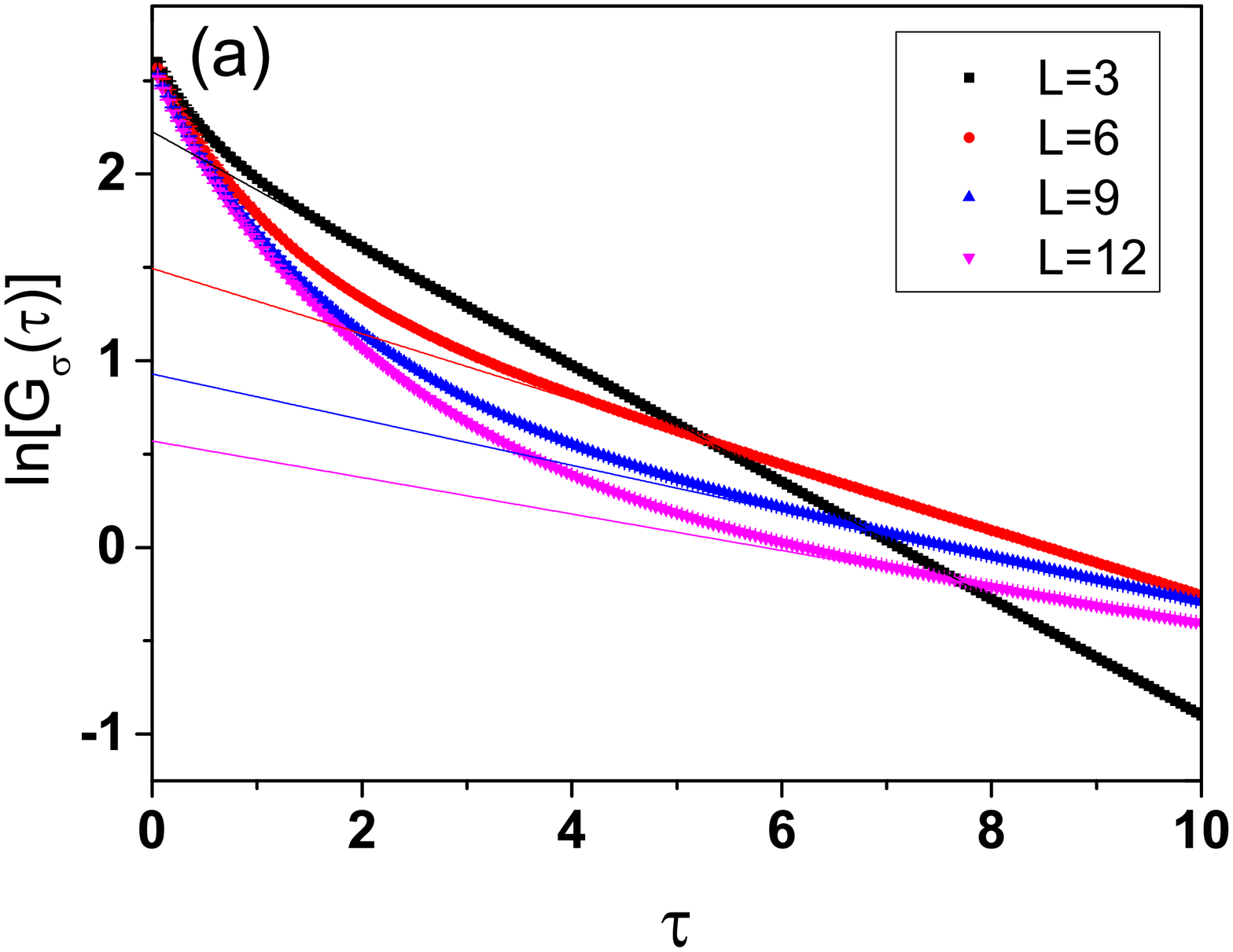}
\includegraphics[width=0.9\columnwidth]{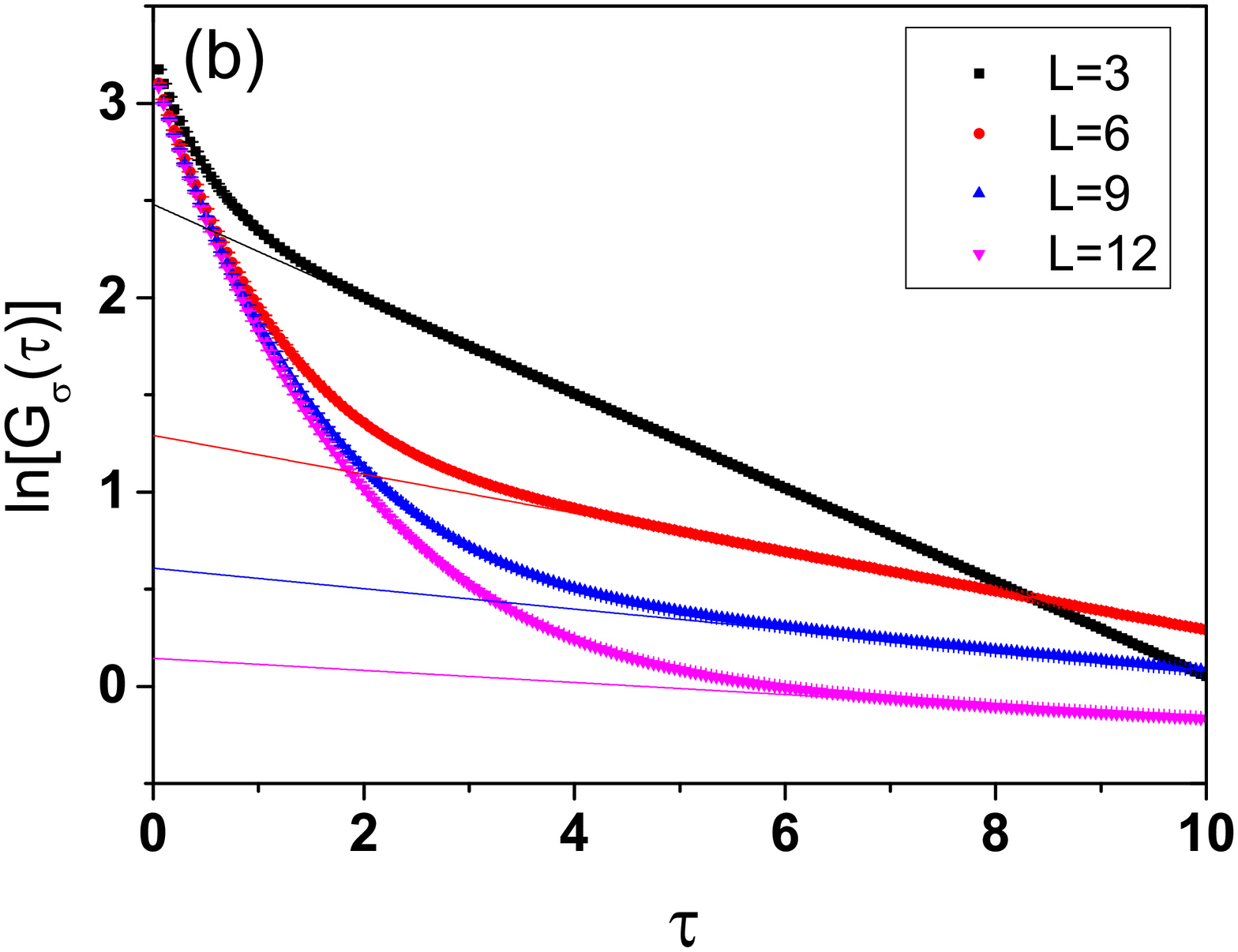}
\caption{$\ln G_{\sigma}(\tau)$ as functions of $\tau$ at $U=6$
for ($a$) SU(4) and ($b$) SU(6) at different values of
lattice size $L$.
Lines are linear fit whose slopes are
the spin gap $\Delta_{\sigma}$.
}
\label{fig:lnGspin}
\end{figure}

\section{Extract the single-particle gap}
\label{append:single_gap}

As discussed in the Sec.~\ref{sec:spgap} of the main text, the
single-particle gap can be obtained from the imaginary-time
displaced Green's function
\bea
G(K,\tau)=\frac{1}{2L^{2}}\sum_{i,j\in A\oplus B}
G(i,j,\tau)e^{i\vec K\cdot (\vec r_{i}-\vec r_{j})}.
\eea
For long enough $\tau$, according to the Lehmann representation,
$G(K,\tau)$ decays in the imaginary time as $\exp(-\Delta_{sp}\tau)$.
We can  extract the single particle gap $\Delta_{sp}$
from the slope of $\ln G(K,\tau)$ vs $\tau$.

In Fig.~\ref{fig:lnGK}, representative plots of $\ln G(K,\tau)$
versus $\tau$ are presented for both the SU(4) and SU(6) cases
at $U=7$ and $U=11$, respectively.
From these plots, we find the linear regime can be easily
achieved without large time-displacement in $\tau$, which
enables us to extract reliable values of $\Delta_{sp}$
shown in the main text.

\section{Extract the spin gap}
\label{append:spingap}

\begin{figure}[htb]
\includegraphics[width=0.9\columnwidth]{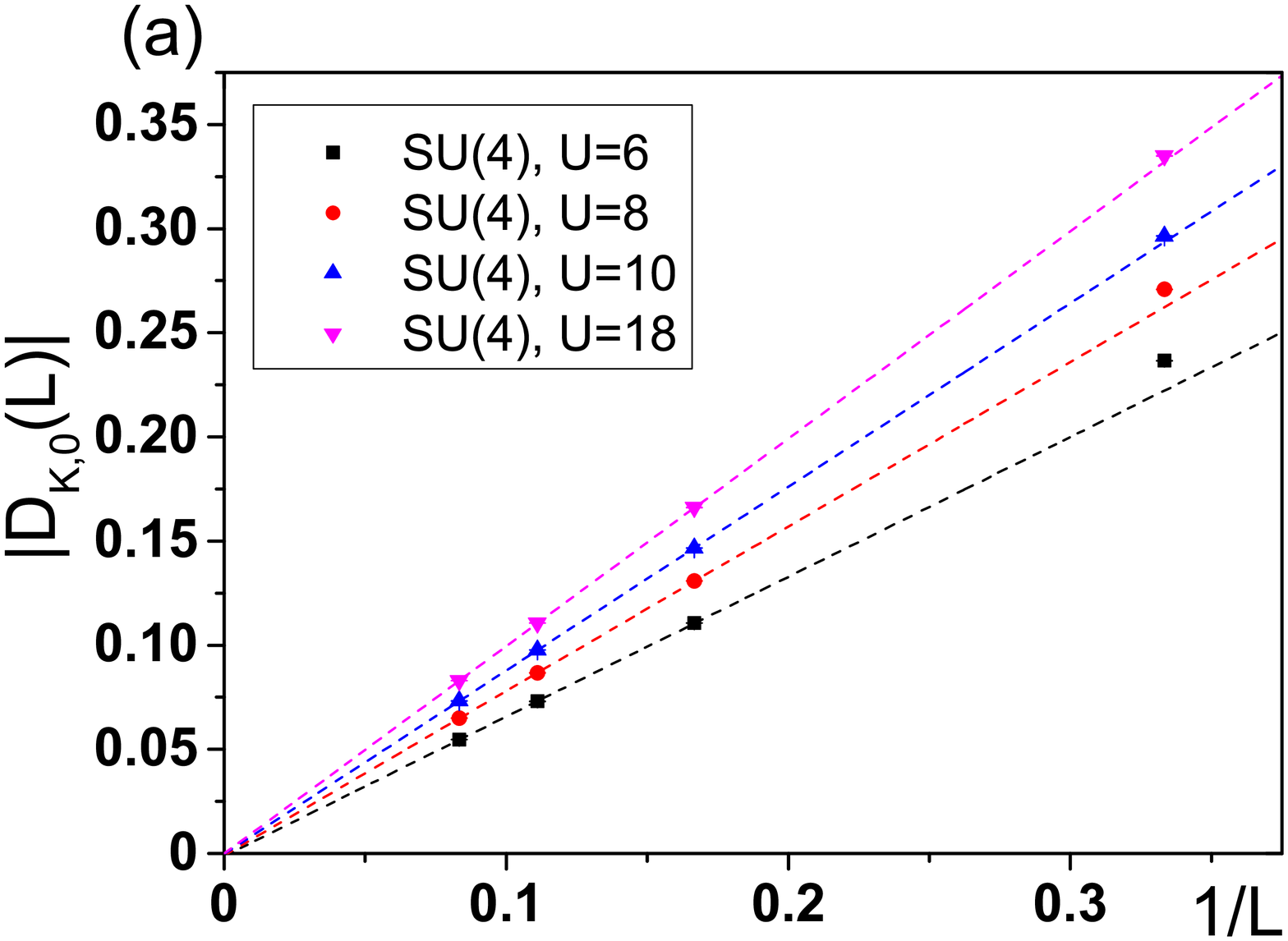}
\includegraphics[width=0.9\columnwidth]{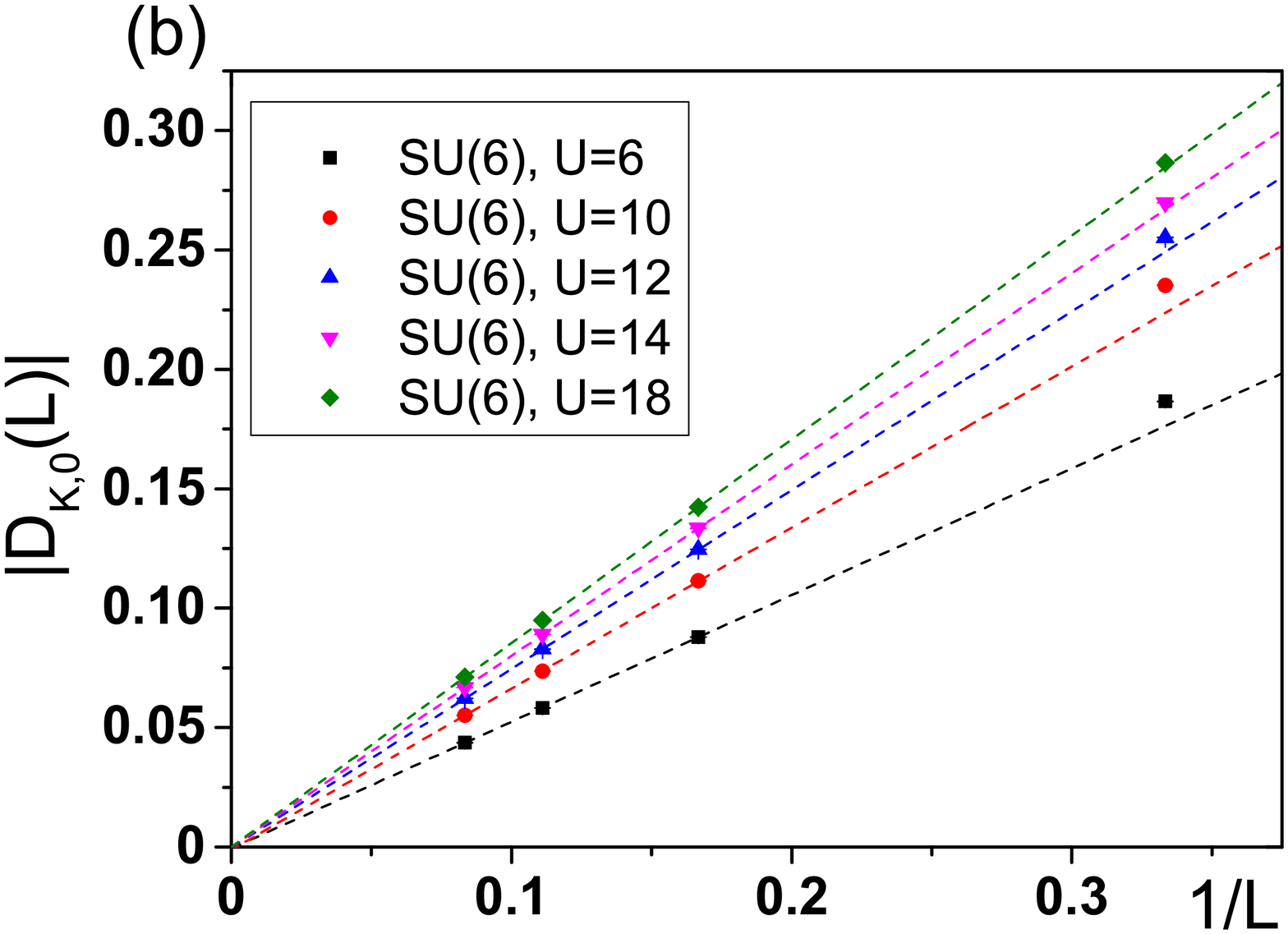}
\caption{Finite size scalings of the trimer order parameter for
different values of $U$ for ($a$) the SU(4) case and ($b$) the SU(6) case.
Linear fit is used starting from $L=6$.
Error bars are smaller than the symbol sizes.
}
\label{fig:trim}
\end{figure}

\begin{figure}[htb]
\includegraphics[width=0.95\columnwidth]{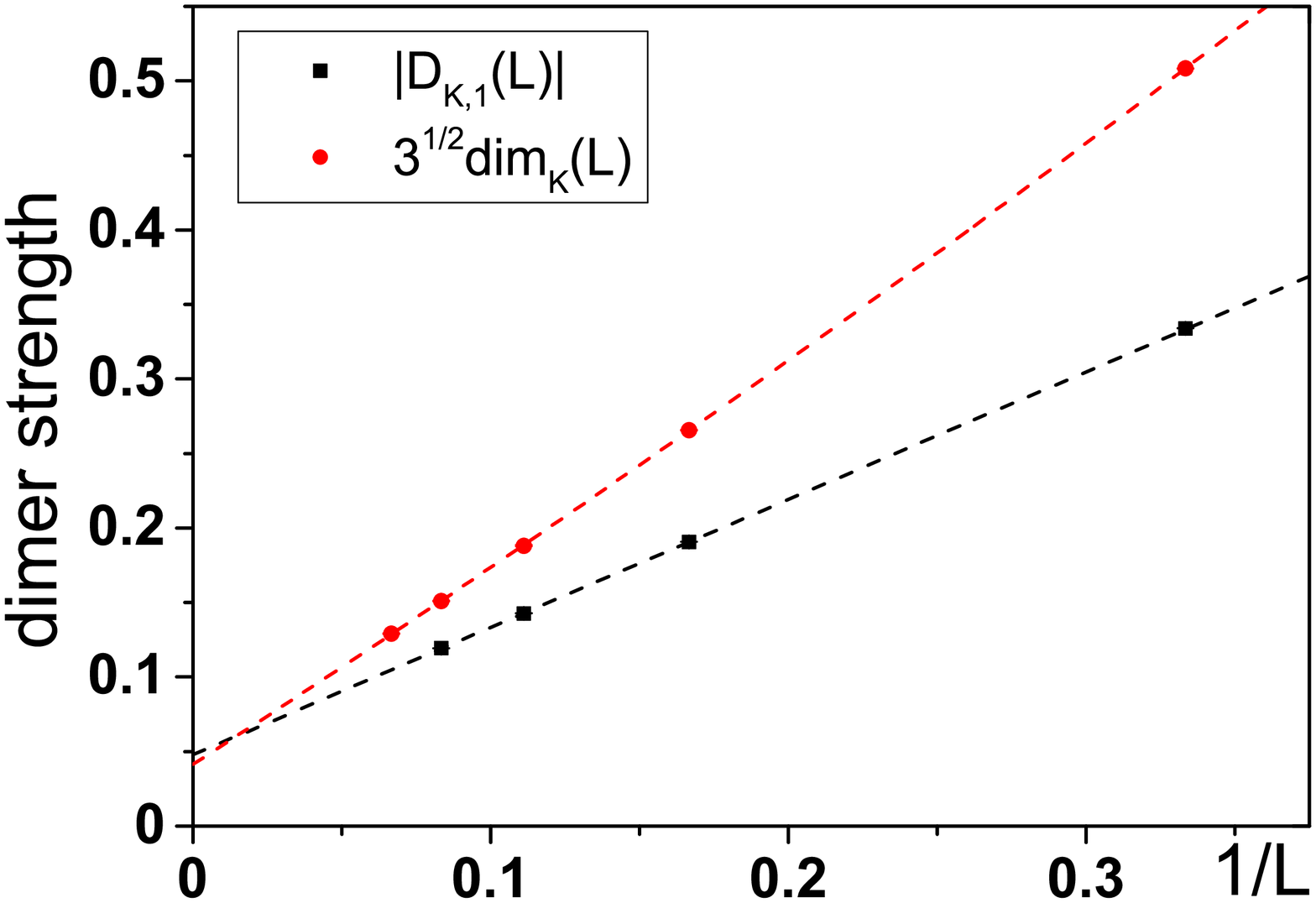}
\caption{A comparison of $|D_{K,1}|$ and $\sqrt{3}\mbox{dim}_{K}$
with parameters $2N=4$ and $U=8$.
Linear fit is used from $L=6$ for $|D_{K,1}(L)|$,
and quadratic fit is used from $L=3$ for $\sqrt{3} \mbox{dim}_{K}(L)$.
}
\label{fig:match}
\end{figure}

As discussed in Sec.~\ref{sect:spin_gap} of the main text, the spin gap
can be extracted from the imaginary time displaced spin-spin correlation
function.
Since the leading spin channel instability occurs at the
N\'{e}el ordering with form factor $(-1)^{i+j}$, we define
the spin-spin correlation function as
\bea
G_{\sigma}(\tau)=\frac{1}{2L^{2}}\sum_{i,j,\alpha,\beta}
\Big\{\left(-1\right)^{i+j}\langle
S_{\alpha\beta,i}(\tau)S_{\beta\alpha,j}(0)\rangle\Big
\}.  \ \ \, \ \ \, \ \ \,
\label{eq:spin_gap_green}
\eea
whose asymptotic behavior at large $\tau$ is
$G_{\sigma}(\tau)\sim \exp(-\Delta_\sigma \tau)$.
Fig.~\ref{fig:lnGspin} shows representative data of $\ln G_{\sigma}(\tau)$
v.s. $\tau$, for both the SU(4) and SU(6) cases, at $U=6$.
The linear regimes at large $\tau$ are used to extract the
spin gap $\Delta_{\sigma}$.

\section{Finite size scaling of trimer and VBS dimer order parameter}
\label{app:dimer}

For simplicity, we have employed Eq. \ref{eq:dimK} to calculate the VBS
dimer order parameter in the main text by only calculating the correlations
among bonds along the same orientations.

\begin{figure}[htb]
\includegraphics[width=0.9\columnwidth]{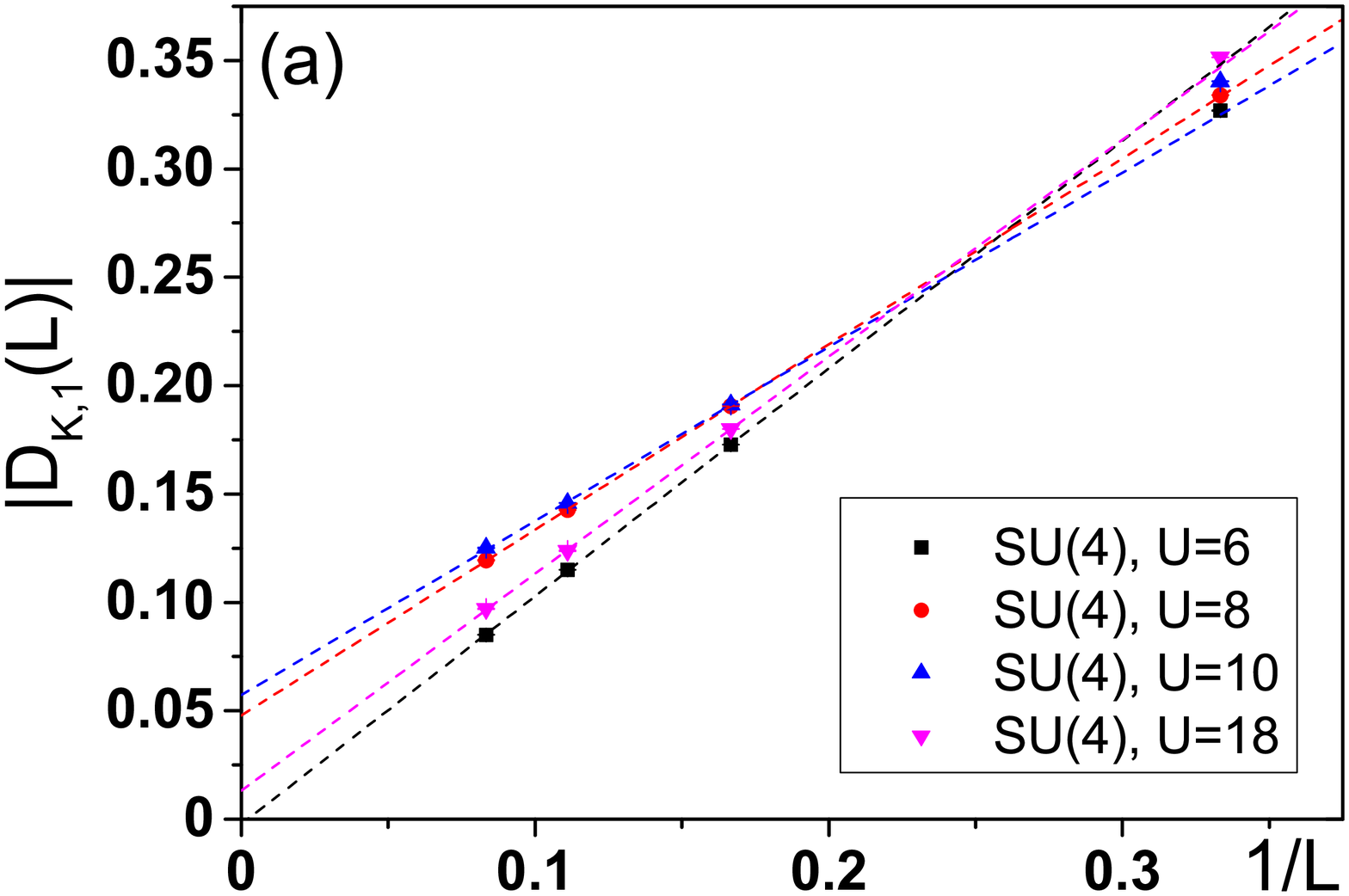}
\includegraphics[width=0.9\columnwidth]{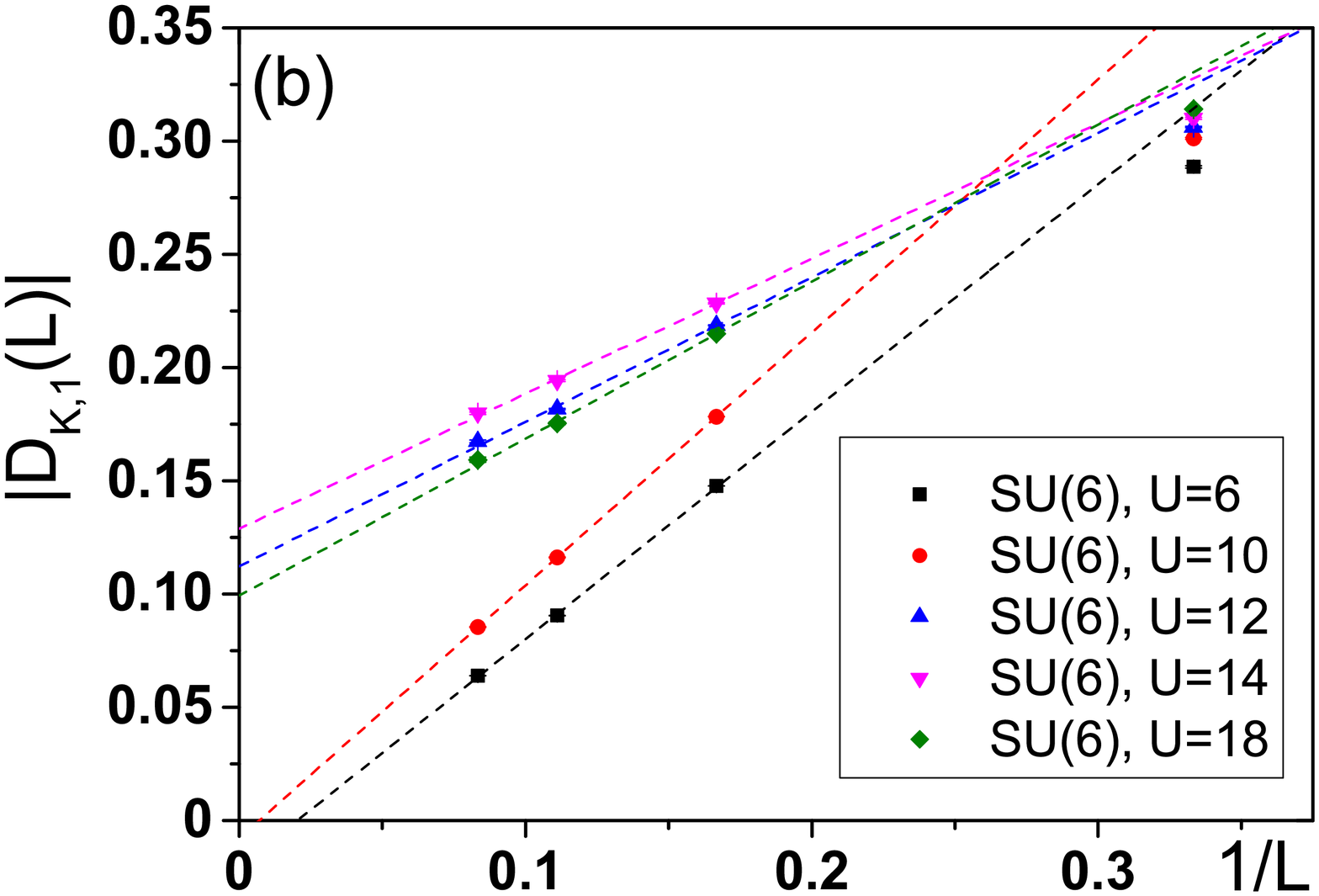}
\caption{Finite size scalings of the VBS dimer parameter $|D_{K,1}|$
at different values of $U$ for ($a$) the SU(4) case and ($b$)
the SU(6) case.
The linear fit is starting from $L=6$, and error bars are
smaller than symbols.
}
\label{fig:DK1}
\end{figure}

We can also use $D_{K,m}$ defined in Eq. \ref{eq:pc_dim} to calculate
the trimer and the VBS dimer order parameters.
The magnitude of the trimer order parameter is defined as
\bea
|D_{K,0}| = \lim_{L\rightarrow \infty} \sqrt{ \langle D^{*}_{K,0}(L)D_{K,0}(L) \rangle },
\eea
and $|D_{K,-1}|$ equals to $|D_{K,0}|$.
Fig.~\ref{fig:trim} shows the finite size scaling of the trimer order
parameter for both SU(4) and SU(6) cases.
Clearly there is no long-range trimer order.
The VBS dimer order parameter can also be calculated from the module
defined as
\bea
|D_{K,1}| = \lim_{L\rightarrow \infty} \sqrt{ \langle D^{*}_{K,1}(L)D_{K,1}(L) \rangle }.
\eea

In the thermodynamic limit, since the trimer correlation is only
short-ranged, the relation between $|D_{K,m}|$ and
$\mbox{dim}_{K}$ is
\bea
|D_{K,1}|=  \sqrt 3 \mbox{dim}_{K}.
\eea
We have compared these two values after the finite-size
scaling for a specific parameter in Fig.~\ref{fig:match}.
They matched well in the thermodynamic limit.
Then in Fig.~\ref{fig:DK1}, the finite size scalings of the VBS dimer
order parameter $|D_{K,1}|$ are presented for different $U$
with both SU(4) and SU(6) cases.
The phase transition point and the non-monotonic
dependence on $U$ are identical comparing with $\mbox{dim}_{K}$.

\begin{figure}[htb]
\includegraphics[width=\columnwidth]{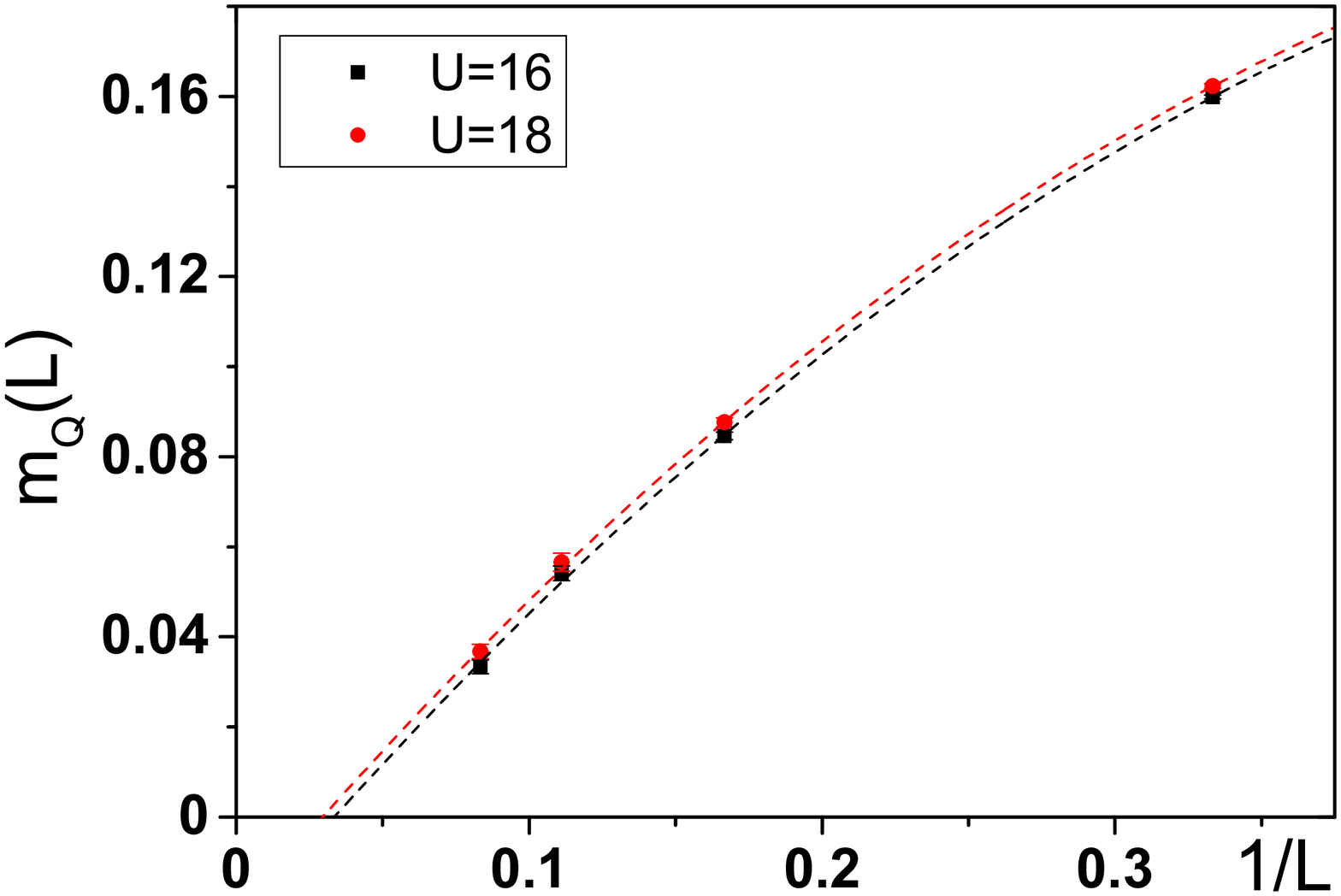}
\caption{The absence of N\'{e}el order $m_Q$ based on the pinning field
method with $U=16$ and $18$ for the SU(4) Hubbard model.
The pinning field is $h=5$. The quadratic polynomial fitting is used,
and error bars are smaller than symbols.
}
\label{fig:neel_pin}
\end{figure}

\begin{figure}[htb]
\includegraphics[width=\columnwidth]{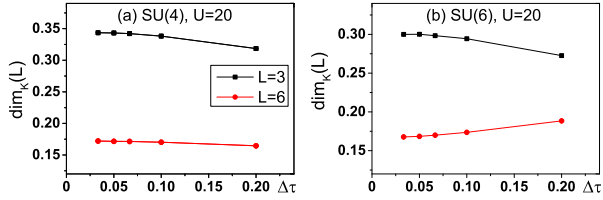}
\caption{The scalings of $\Delta\tau$ for the VBS dimer order
$\mbox{dim}_K$. Error bars of QMC data are smaller than symbols. }
\label{fig:tau_dimer}
\end{figure}

\begin{figure}[htb]
\includegraphics[width=\columnwidth]{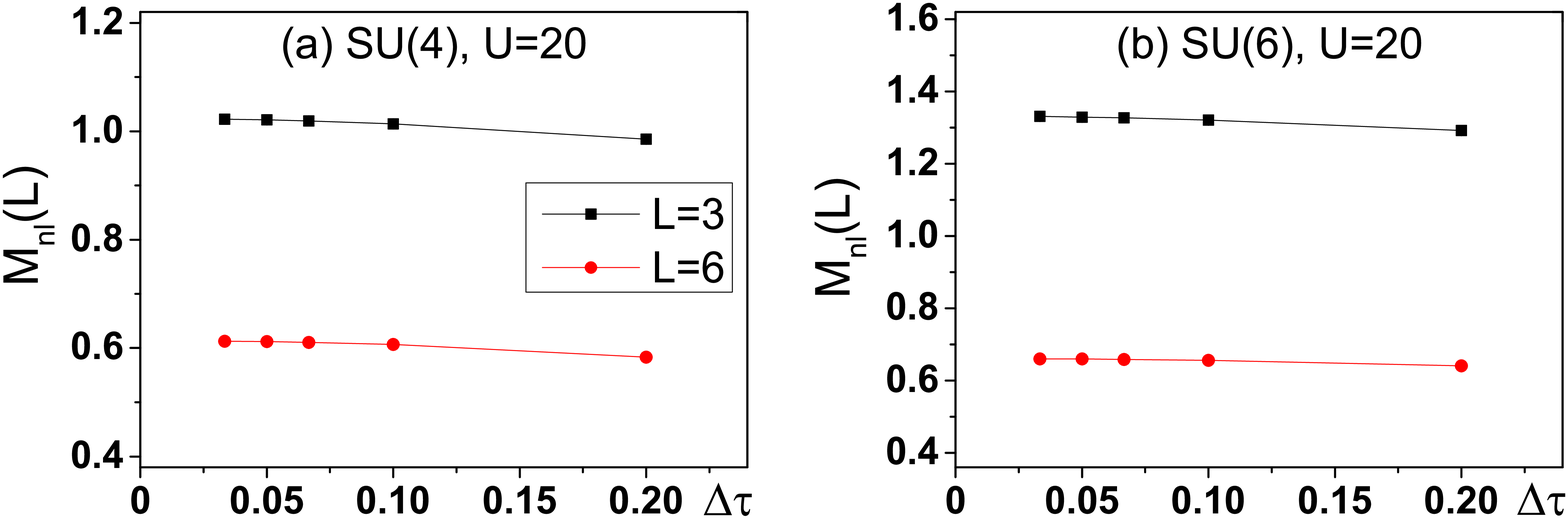}
\caption{The scalings of $\Delta\tau$ for the N\'{e}el order $M_{nl}$
based on the structure factor.
Error bars of QMC data are smaller than symbols.}
\label{fig:tau_neel}
\end{figure}

\section{Using pinning field method to rule out N\'{e}el order}
\label{app:pin}

To further clarify whether there is a weak long-range N\'{e}el order at
large $U$ in the SU(4) case, we employ the pinning field method.
Due to its sensitivity to weak N\'{e}el order, this method has been
applied to the SU(2) Hubbard model in the honeycomb lattice
\cite{Assaad2013} and the SU($2N$) Hubbard model in the
square lattice \cite{Wang2014}.
Following the procedure in Ref. [\onlinecite{Wang2014}], a pinning
term is added into the original Hamiltonian Eq. \ref{eq:hubbard} as
\bea
H_{pin}(i_0,j_0)= h \left[ m(i_0) -m (j_0) \right],
\label{eq:bin}
\eea
in which the onsite magnetic moment is defined as
$m(i)=\sum_{\alpha=1}^N n_\alpha(i)
-\sum_{\alpha=N+1}^{2N} n_\alpha(i)$,
and $i_{0}$ and $j_{0}$ are two nearest-neighbor sites in a unit cell.
Since the SU($2N$) symmetry is explicitly broken, the N\'{e}el order
is induced by the external field.
Since the pinning field only applies on two nearest-neighboring sites,
and the symmetry breaking effect is at the order of $1/L$.
The long-range N\'{e}el order is defined as
\bea
m_{Q}&=&\lim_{L\to \infty}\frac{1}{2L^2}\sum_i (-1)^i m(i).
\eea
There is only an overall factor difference between $M_{nl}$ in main
text and $m_Q$ here in the thermodynamic limit if N\'{e}el order
actually exists.

We have simulated the N\'{e}el ordering for the SU(4) Hubbard model
at $U=16$ and $18$ using the pinning field method.
$\Delta\tau$ is set as 0.05 after the discrete $\Delta\tau$
scaling.
For the pinning field method, much longer projection time
is needed because the pinning fields break the SU($2N$)
symmetry and thus cause much smaller gap. \cite{Assaad2013,Wang2014}
We also performed the scaling over $\beta$, and find that very large $\beta$
(as long as $350$) for $L=12$ is needed to ensure the convergence.
The simulation results  are presented
in Fig. \ref{fig:neel_pin}, which show the absence of the
long-range N\'{e}el order for $U=16$ and $18$ in the SU(4) case
in consistent with the results in the main text.

\section{The discrete $\Delta\tau$ scaling}
\label{app:tau}

In this section, we examine the dependence of the QMC results on the
imaginary time discretization $\Delta\tau$.
Since the discretization error coming from the second-order Suzuki-Trotter
decomposition is of order $(\Delta\tau)^3U^2t$, which becomes the most
severe when $U$ increases, we only need to check our QMC results
at the largest $U=20$ in our simulations.

\begin{figure}[htb]
\includegraphics[width=\columnwidth]{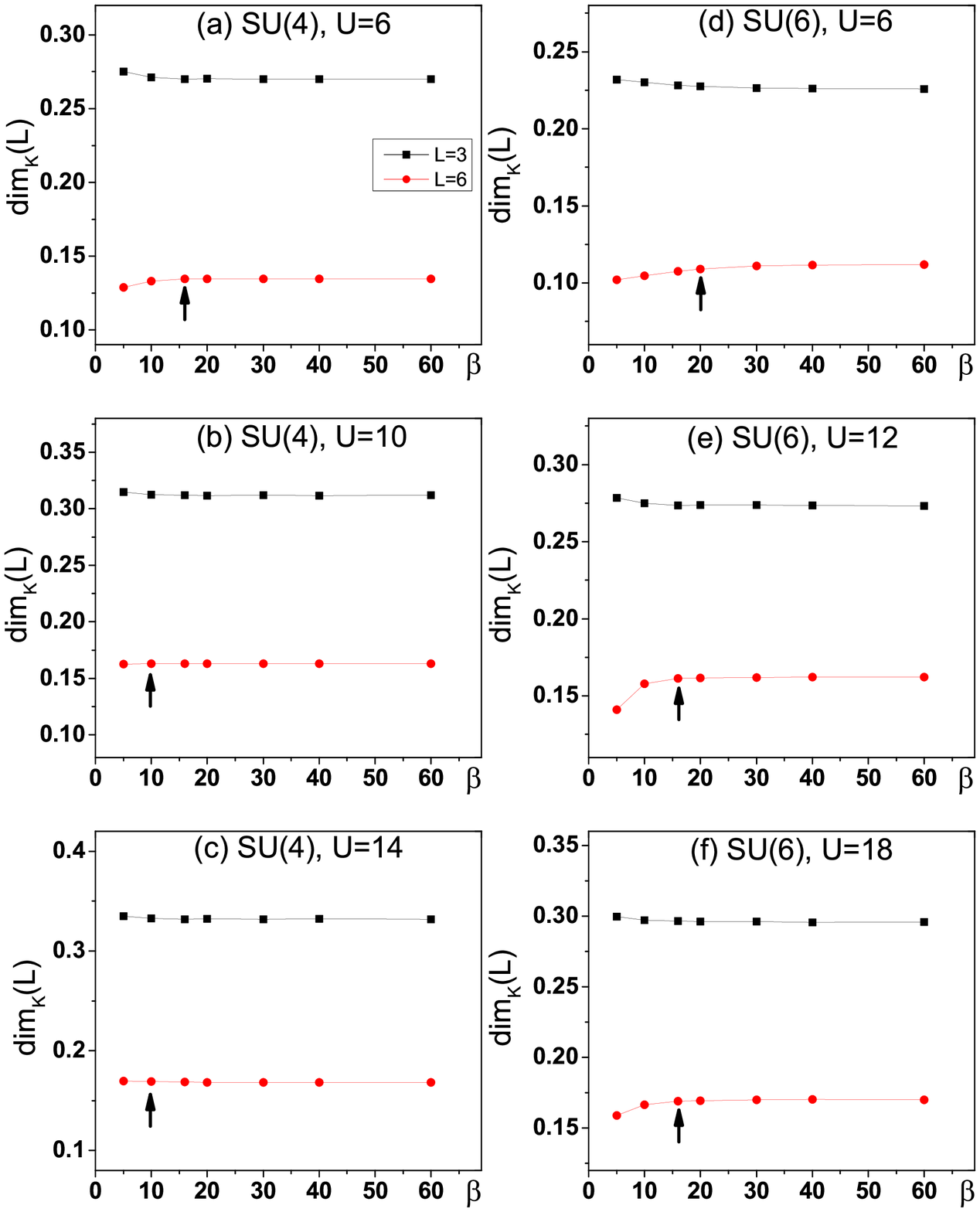}
\caption{The scalings of the projection time $\beta$ for the dimer VBS
order $\mbox{dim}_K$.
Arrows denote the $\beta$ at which the results converge.
Error bars of QMC data are smaller than symbols. }
\label{fig:beta_dimer}
\end{figure}

We first present the $\Delta\tau$ scalings for the VBS dimer order with
$U=20$ for the SU(4) and SU(6) cases in Fig.~\ref{fig:tau_dimer}
($a$) and ($b$), respectively.
The data show that $\Delta\tau=0.05$ is small enough to ensure the
convergence, regardless of the lattice size.
In Fig. \ref{fig:tau_neel} ($a$) and ($b$), we further present
the $\Delta\tau$ scalings for the N\'{e}el order
for the SU(4) and SU(6) cases, which also show
convergence at $\Delta\tau=0.05$.
Based on these scalings, we use $\Delta\tau=0.05$ when performing all
the QMC simulations in the main text.

\section{The finite $\beta$ scaling}
\label{app:beta}

The projection time $\beta$ needs to be long enough to arrive at the
ground state.
For a finite size system with a linear size $L$, supposing the first
excitation gap is $\Delta (L)$, $\exp(-\beta\Delta)\ll 1$ is required.
$\Delta(L)$ often scales as $1/L$ with antiferromagnetic correlations.
If the system is gapped at $L\to \infty$, $\Delta(L)$ drops to a
finite value at sufficiently large values of $L$.
Therefore, we try $\beta$ proportional to $L$.

\begin{figure}[htb]
\includegraphics[width=\columnwidth]{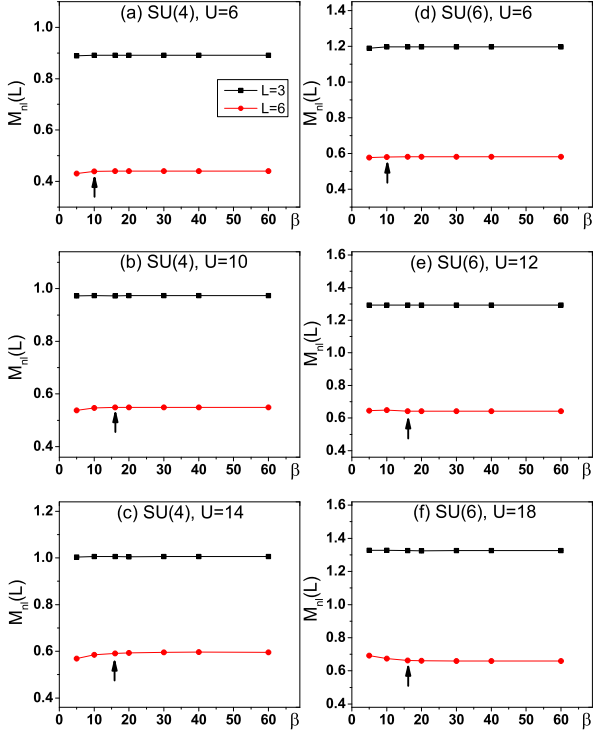}
\caption{The scalings of the projection time $\beta$ for the N\'{e}el
order $M_{nl}$.
Arrows denote the $\beta$ at which the results converge.
Error bars of QMC data are smaller than symbols.}
\label{fig:beta_neel}
\end{figure}

In Fig. \ref{fig:beta_dimer} ($a$)- ($f$), we present the scalings of
$\beta$ for the VBS dimer order parameter for both the SU(4) and SU(6) cases.
In each case, three different interactions $U$ are chosen.
From the scaling results, we find that $\beta=10 \sim 16$ is enough
for convergence for $L=6$ in most cases, as denoted by the arrows.
Therefore, we choose $\beta=40$ in the simulations to ensure the
projection time is long enough to reach the ground states
for $L$ up to 15.
Similar analyses are performed on the $\beta$-dependence of
the N\'{e}el order parameter in Fig.~\ref{fig:beta_neel}
($a$) - ($f$).
We keep choosing the above projection time $\beta=40$
in all our QMC simulations.

\bibliography{sundirac,sun_honey,spin32}

\end{document}